\makeatletter
\@ifundefined{@parse@version@dash}{%
\def\@parse@version#1{\@parse@version@0#1}
\def\@parse@version@#1/#2/#3#4#5\@nil{%
\@parse@version@dash#1-#2-#3#4\@nil}
\def\@parse@version@dash#1-#2-#3#4#5\@nil{%
  \if\relax#2\relax\else#1\fi#2#3#4 }
}{}
\makeatother

\documentclass[twocolumn,secnumarabic,amsmath,amssymb,aps,pra,superscriptaddress,eqsecnum,longbibliography]{revtex4-2}
\usepackage{graphicx}
\usepackage{bm}
\usepackage{xcolor}
\usepackage[normalem]{ulem}
\usepackage{braket}
\usepackage{algorithmic}
\usepackage{algorithm}
\usepackage{booktabs}
\usepackage{physics}
\usepackage{soul}
\usepackage{tikz-imagelabels}
\usepackage{latexsym}

\usepackage{amsthm}
\theoremstyle{definition}

\newtheoremstyle{mystyle}%
    {}
    {}%
    {\itshape}%
    {}%
    {\bfseries}%
    {.}%
    { }%
    {\thmname{#1}\thmnumber{ #2}\thmnote{ (#3)}}
\theoremstyle{mystyle}

\usepackage[,
colorlinks=true,
urlcolor=blue,
citecolor=blue,
linkcolor=blue,
hyperfootnotes=false]{hyperref}

\begin{document}
\title{Variational quantum eigensolver with embedded entanglement \\ using a tensor network ansatz}

\author{Ryo Watanabe}
\affiliation{Graduate School of Engineering Science, Osaka University, 1-3 Machikaneyama, Toyonaka, Osaka 560-8531,
Japan}

\author{Keisuke Fujii}
\affiliation{Graduate School of Engineering Science, Osaka University, 1-3 Machikaneyama, Toyonaka, Osaka 560-8531,
Japan}
\affiliation{Center for Quantum Information and Quantum Biology, Osaka University, Toyonaka, Osaka 560-0043, Japan}
\affiliation{RIKEN Center for Quantum Computing (RQC), Wako, Saitama 351-0198, Japan}

\author{Hiroshi Ueda}
\affiliation{Center for Quantum Information and Quantum Biology, Osaka University, Toyonaka, Osaka 560-0043, Japan}
\affiliation{Computational Materials Science Research Team, RIKEN Center for Computational Science (R-CCS), Kobe, Hyogo 650-0047, Japan}

\begin{abstract}
In this study, we introduce a tensor network (TN) scheme into the entanglement augmentation process of the synergistic optimization framework [arXiv: 2208.13673] by Rudolph et al. to build its process systematically for inhomogeneous systems.
Our synergistic approach first embeds the variational optimal solution of the TN state with the entropic area law, which can be perfectly optimized in conventional (classical) computers, in a quantum variational circuit ansatz inspired by the TN state with the entropic volume law. Next, the framework performs a variational quantum eigensolver (VQE) process with embedded states as the initial state.
We applied the synergistic to the ground state analysis of the all-to-all coupled random transverse-field Ising/XYZ/Heisenberg model, employing the binary multi-scale entanglement renormalization ansatz (MERA) state and branching MERA states as TN states with entropic area law and volume law, respectively.
We then show that the synergistic accelerates VQE calculations in the three models without an initial parameter guess of the branching-MERA-inspired ansatz and can avoid a local solution trapped by a standard VQE with the ansatz in the Ising model.
The improvement of optimizers for MERA in all-to-all coupled inhomogeneous systems, enhancement, and potential synergistic applications are also discussed.

\end{abstract}

\date{\today}

\maketitle

\section{Introduction}\label{sec:Intro}
The ground state of quantum many-body systems is essential to understand quantum phenomena in many-body physics.
However, solving quantum many-body problems is generally difficult with conventional (classical) computers because they require the treatment of Hilbert spaces that grow exponentially with an increasing system size of $N$.

Using quantum computers has been attracting attention to address the problems' limitations.
Actually, quantum algorithms such as the quantum phase estimation~\cite{Daskin_2018}, the qubitization~\cite{Low_2019}, and the quantum singular value transformation~\cite{Gily_n_2019} with quantum acceleration, unable us to access the ground state for large quantum many-body systems that intractable for classical computers.

However, current quantum computers are noisy intermediate-scale quantum (NISQ) devices~\cite{Preskill2018quantumcomputingin} that suffer from errors and are limited in size.
Quantum algorithms that can be executed using shallow quantum circuits are required to make them more practical.
In this context, one of the NISQ-aware methods developed is the Variational Quantum Algorithm (VQA)~\cite{Cerezo2021}.
This method involves constructing variational models through parameterized quantum circuits (also called ansatz) on a quantum computer and optimizing the cost function using classical computers.
VQAs, specifically the variational quantum eigensolver (VQE), have been applied to search for the ground state of quantum systems~\cite {Peruzzo2014}.

To obtain meaningful results using VQAs on NISQ, the ansatz utilized to represent the target state with the required precision should have few internal parameters and be as shallow as possible concerning the circuit depth.
This condition is crucial for mitigating the noise effects inherent to NISQ devices and preventing the occurrence of the barren plateau (BP) problem~\cite{McClean2018}, which is associated with an overly excessive variational space.
Although the BP problem can be circumvented by starting the calculation from appropriate initial parameters~\cite{Grant2019initialization}, it is generally nontrivial to set such parameters.
Therefore, a systematic procedure for setting up a high-quality ansatz is desired.

As a strategy to satisfy this requirement, the procedures of putting more classical computer resources into a series of processes of VQE are currently attracting attention~\cite{okada2022identification, arxiv.2208.13673}.
The VQE with a synergistic framework~\cite{arxiv.2208.13673} is a quantum-classical hybrid algorithm that has received much attention to avoid BP problems and works efficiently with less use of NISQ.
The synergistic framework has two key concepts: sophisticated variational methods on classical computers ideally provide the initial state for the VQE, and quantum gates added during the VQE calculation and the quantum circuit in which the classically optimized solution is efficiently embedded have different topologies.
The second concept is referred to as the entanglement augmentation process.

The use of the tensor network (TN) method~\cite{ORUS2014117,Orus2014,ran2020tensor,ONU2022TN} for the first concept of the synergy framework is a natural idea because the TN method for quantum states represents the wave function as a contraction of many tensors with small degrees of freedom (TN representation) and is a way to realize the variational method of large quantum many-body systems on a classical computer; there is an ongoing effort to convert TN states into quantum circuit representations today~\cite{Liu_PRR2019, PhysRevX.12.011047,miao2023convergence}.

Actually, in Ref.~\cite{arxiv.2208.13673}, the authors used the matrix product state (MPS)~\cite{PhysRevLett.69.2863,SCHOLLWOCK201196,PhysRevB.97.045145}, which has a one-dimensional (1D) topology, as a variational method on classical computers, and employed all-to-all topologies of fully parameterized SU(4) gates for the augmentation process.
However, as mentioned in Ref~\cite{arxiv.2208.13673}, constructing an all-to-all topology is a nontrivial task and is likely not scalable on NISQ devices.
Therefore, to achieve a synergistic framework irrespective of the task that can be implemented on near-to-mid-term quantum computers, it is necessary to propose a systematic entanglement augmentation process that can qualitatively and dramatically change the entanglement structure of the ansatz by only adding very sparsely coupled quantum gates.

In this paper, we introduce the concept of a TN structure into this entanglement augmentation process and propose a procedure to efficiently and systematically augment the entanglement representation capability of a quantum circuit ansatz.
Specifically, we focus on embedding a TN state with the entropic area law~\cite{RevModPhys.82.277} after variational calculation on a classical computer in a portion of the entanglement-augmented quantum circuit ansatz, which is equivalent to a TN state with the entropic volume law.
This entangled embedding VQE (EEVQE) approach can accelerate the calculations of VQE with a TN-inspired ansatz with an entropic-volume law. Discussions of embedding focused on TN with the entropic volume law and its connection to VQE, which is an attempt not seen in pioneering works~\cite{Goldsborough_PRB2017,miao2023convergence,Grant2018} related to our study.
Investigations of the TN state with the entropic volume law are essential for analyzing quantum states in complex problems, such as quantum chemical systems, long-range interacting systems, random spin systems, and nonlinear problems~\cite{PhysRevA.101.010301} which involve non-uniform interactions among all qubits, and simulations of real-time evolution for quantum systems because the quantum states may not satisfy the entropic area law in these systems.
For specific numerical verification of EEVQE, we employ a 1D binary multi-scale entanglement renormalization ansatz (MERA)~\cite{Vidal_2008, Evenbly_Vidal_PhysRevB.79.144108} for variational methods on a classical computer and use branching MERA networks~\cite{Evenbly_2014,PhysRevB.89.235113} as the quantum circuit ansatz after entanglement augmentation.
The target Hamiltonians are all-to-all coupled random transverse-field Ising/XYZ/Heisenberg models whose ground states can violate the entropic area law.
As we expect, through the EEVQE process, we confirm for three models that TN states after the variational optimization on classical computers can be further improved in terms of variational energy while benefiting from the initialization-problem-free and accelerating VQE calculations.
Furthermore, we also confirm that EEVQE can avoid traps in locally optimal solutions (or broadly BP) in the random transverse field Ising model.

As part of our efforts to advance the synergistic framework, we compared optimizers to update MERA for all-to-all coupled random systems.
This mainly includes the Evenbly-Vidal method~\cite{Evenbly_Vidal_PhysRevB.79.144108}, the standard in MERA optimization, its improvements developed in this work (refer to Sec.~\ref{sec:improved procedures}), and the Broyden-Fletcher-Goldfarb-Shanno (BFGS) method, which is the standard in parameterized quantum circuit ansatz optimization.
Our comparison results showed that the BFGS method was the most effective in reducing the variational energy for all the studied models.
This result indicates the importance of introducing the knowledge of optimizers used in VQE to accelerate the computation of variational methods with TN on the classical computer in the application of the synergistic framework to investigate low-energy states of quantum chemical calculations, which is the most desired application of VQE for all-coupled inhomogeneous systems.

The remainder of this paper is organized as follows: in Sec.~\ref{sec:reviews}, we review the basics of the components of EEVQE, namely the TN structures of MERA and branching MERA, Evenbly-Vidal sequential optimization method, VQE, and the synergistic framework; in Sec.~\ref{sec:EEVQE}, we describe our propositions, i.e., the procedure of embedding MERA states into a quantum circuit ansatz inspired by branching MERA states and the procedure of a modified Evenbly-Vidal method; in Sec.~\ref{sec:numerical_sim}, after investigating the performance of optimizers for MERA states in a specific set of all-to-all coupled models, we present the results of a performance study of EEVQE for the models.
Finally, in Sec.~\ref{sec:summary}, we summarize the study and discuss future studies on EEVQE.

\section{Reviews for components of EEVQE}\label{sec:reviews}
In this section, we review the fundamentals of the components of EEVQE: the TN structures of 1D binary MERA and branching MERA, optimization methods for MERA, the calculation procedure of VQE, and the synergistic framework.
One familiar with these components may skip to the next section, which discusses how to embed the optimized MERA state in branching-MERA-inspired quantum circuits, and the scrutiny of optimization methods for MERA in the all-to-all coupled model.

\subsection{TN structures of 1D binary MERA and branching MERA}\label{subsec:tn_structure}
Here, let us consider the 1D binary MERA, whose schematic diagram is depicted in Fig.~\ref{fig:binaryMERA, isometric_condition}(a).
\begin{figure}[!h]
    \begin{center}
    \includegraphics[clip, width=3.3in]{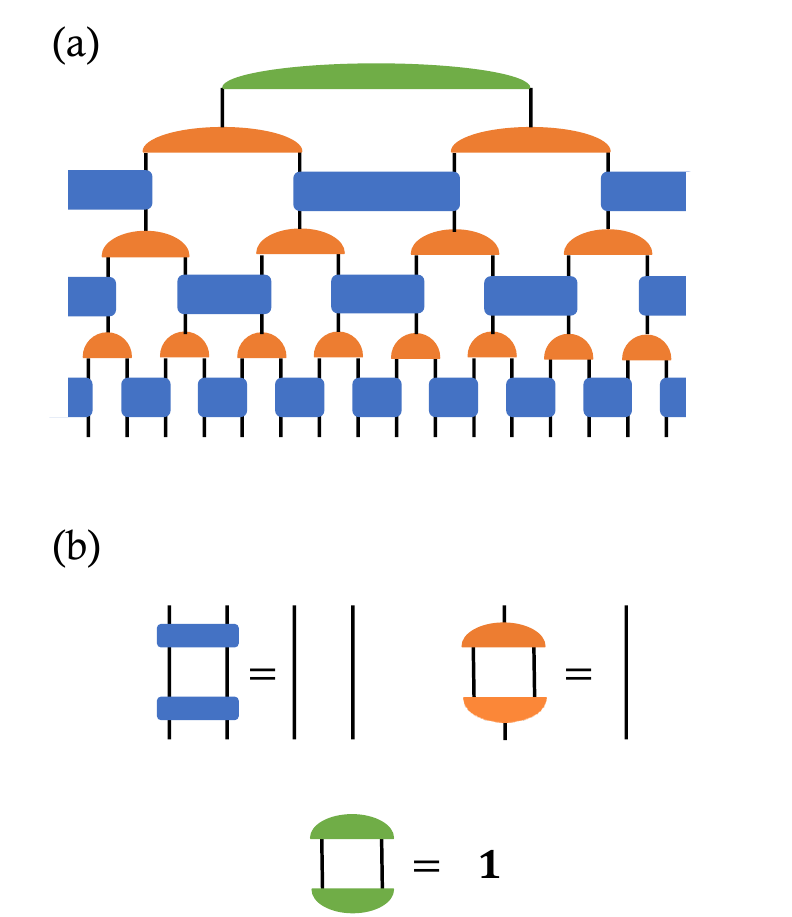}
    \caption{Schematic diagrams of (a) the binary MERA for the 16-site system and (b) isometric conditions of disentangler, isometry, and top tensors.}
    \label{fig:binaryMERA, isometric_condition}
    \end{center}
\end{figure}
The binary MERA network for finite-size systems consists of fourth-, third-, and second-order tensors, called disentanglers $u$, isometries $w$, and top tensors $t$ (see Fig.~\ref{fig:binaryMERA, isometric_condition}).
In the following, to simplify the discussion, the degrees of freedom (bond dimension) of each leg of all tensors are assumed to be the same positive integer $\chi$ (see the original paper in MERA~\cite{Evenbly_Vidal_PhysRevB.79.144108} to extend the discussion to the case where the bond dimension is tensor-dependent).
Each $u$, $w$, and $t$ on the network always satisfies the following orthonormality or isometric conditions
\begin{eqnarray}
\sum_{cd} \left(u^{ab}_{cd}\right)^* u^{a'b'}_{cd} & = & \delta_{aa'} \delta_{bb'}, \label{eq:isometric_u_1}\\
\sum_{ab} \left(u^{ab}_{cd}\right)^* u^{ab}_{c'd'} & = & \delta_{cc'} \delta_{dd'}, \label{eq:isometric_u_2}\\
\sum_{ab} \left(w^{c}_{ab}\right)^* w^{c'}_{ab} & = & \delta_{cc'}, \label{eq:isometric_w}\\
\sum_{ab} \left(t_{ab}\right)^* t_{ab} & = & 1~,
\end{eqnarray}
where $\delta_{\alpha\alpha'}$ with $\alpha \in \{a,b,c,d\}$ is the Kronecker delta.
These conditions are shown schematically in Fig.~\ref{fig:binaryMERA, isometric_condition}(b).
Owing to these conditions, the quantum state $\ket{\Psi}$ represented by MERA is always normalized, that is, $\braket{\Psi}=1$.
This study considers a position-dependent TN, where each tensor has different tensor elements depending on its placement in the network.

\begin{figure}[!h]
    \centering
    \includegraphics[clip, width=3.0in]{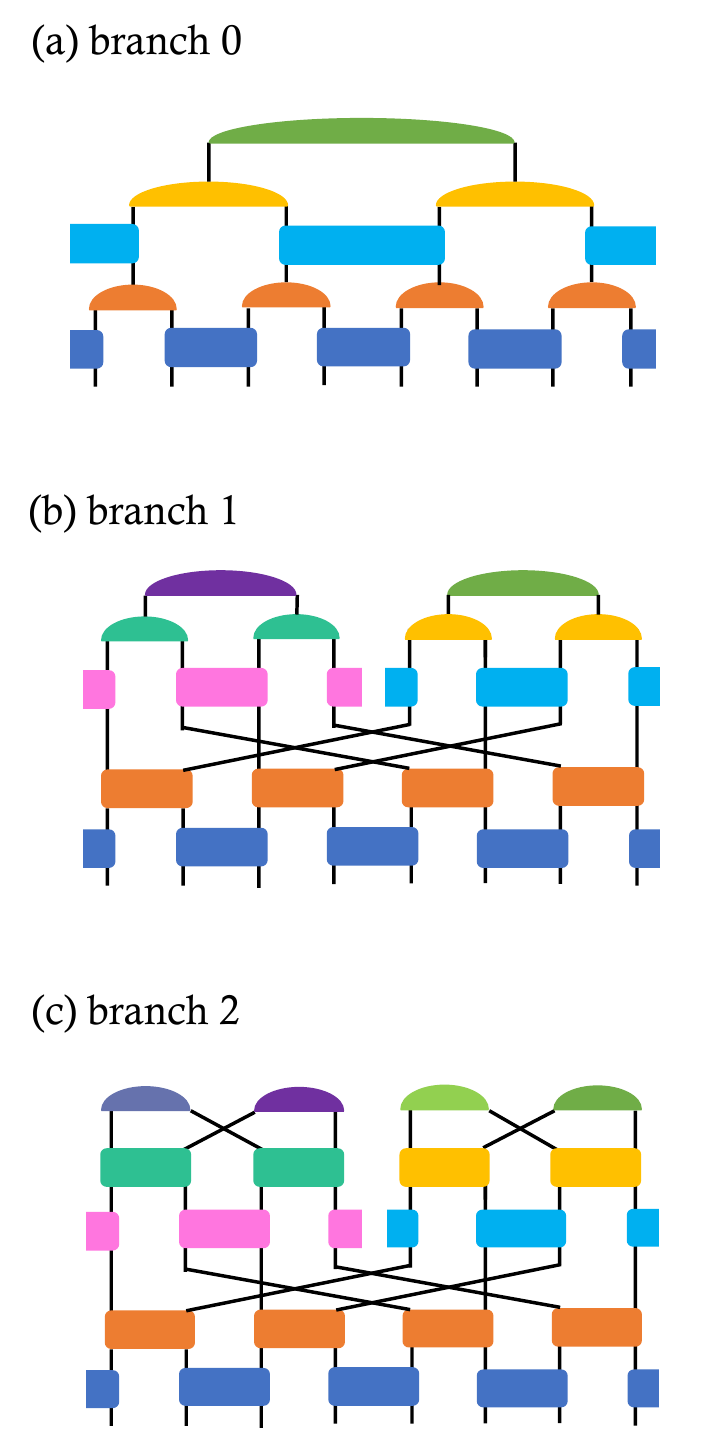}
    \caption{Pattern diagram of the binary MERA to the branching MERA in an 8-site system. Since branches can be added at each layer, we name the diagram to branch0, branch1, branch2,$\dots$ for each number of branches.}
    \label{fig:branching}
\end{figure}

To achieve a more expansive variational space, the 1D branching binary MERA in Fig.~\ref{fig:branching} introduces a branching structure into isometry while keeping layer structures of the MERA.
In the case of the branching binary MERA, the orthonormality of the isometry in which the bifurcation is introduced is identical to that of the disentangler $u$.
Since the branching structure can be arbitrarily introduced in each isometry layer in the binary MERA as shown in Fig.~\ref{fig:branching}, variations of $2^{n-1}$-pattern branching structures are possible in a $2^n$-site system.
This paper focuses only on a full-branching MERA, which introduces branching in all isometry layer , for example, branch 2 in
Fig.~\ref{fig:branching} because only the full-branching MERA satisfies the entropic volume law~\cite{PhysRevA.101.010301}.
Of course, the more branching we adopt, the larger Hilbert space the branching MERA can search so we may consider branching patterns depending on the target system.

\subsection{Optimization scheme for MERA}\label{subsec:tn_opt}
\begin{figure}[!h]
    \begin{center}
    \includegraphics[clip, width=3.3in]{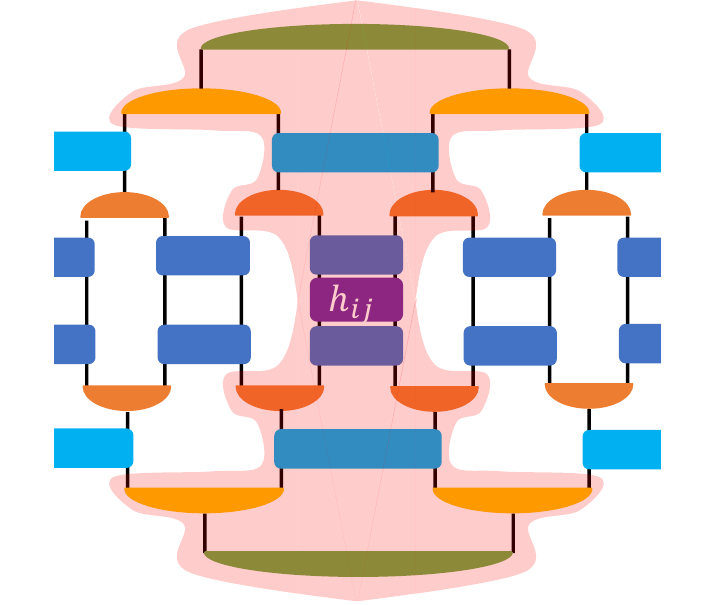}
    \caption{Schematic diagram of $e_{ij}$ with $(i,j)=4,5$, and the light-shaded region shows the causal cone structures. Diagrams outside the causal cone are trivially simplified to the identity operator by using isometric conditions.}
    \label{fig:causal_cone}
    \end{center}
\end{figure}

This subsection presents the Evenbly-Vidal algorithm, which is used as the standard optimization of MERA.
Given the Hamiltonian $\mathcal{H}$ of the targeted system, the variational energy $E$ can be evaluated as $\ev{\mathcal{H}}{\Psi}$, where $\ket{\Psi}$ is the variational state in terms of the binary MERA.
We now consider that the Hamiltonian consists of the sum of the two-body interactions $h_{ij}$ between $i$- and $j$-th sites, namely $\mathcal{H}=\sum_{i<j} h_{ij}$.
In this case, the expectation value of each interaction $e_{ij}=\ev{h_{ij}}{\Psi}$can be summed to obtain the variational energy $E$.
For practical calculations, we should perform the minimum contraction, which depends on the $(i,j)$-pairs by utilizing the isometric condition of $u,w,t$, or causal cone as shown in Fig~\ref{fig:causal_cone}, and the typical computational cost of evaluating $e_{ij}$ is known to be $O(\chi^9 \log N)$ if the system size is sufficiently large ~\cite{Evenbly_Vidal_PhysRevB.79.144108}.
For example, given a MERA, suppose we want to optimize an isometry $w$ while keeping the rest of the tensors fixed.
The energy $E$ depends quadratically on $w$ and $w^{\dag}$, namely
\begin{equation}
    E(w) = \sum_{abca'b'c'}\sum_{\langle i<j \rangle} \left(w^{c}_{ab}\right)^* \left[\mathcal{E}_{ij}\right]^{abc}_{a'b'c'} w^{c'}_{a'b'} + C
\end{equation}
where $\langle i<j \rangle$ refers to the set of site pairs specifying the two-body pair interaction $h_{ij}$ that contributes to the optimization of $w$, and $\mathcal{E}_{ij}$ is the environment tensor obtained by hollowing out the diagrams corresponding to $w$ and $w^{\dag}$ from the causal cone necessary to evaluate $e_{ij}$~\cite{Evenbly_Vidal_PhysRevB.79.144108}, and $C=\sum_{\overline{\langle i<j\rangle}} e_{ij}$ with $\sum_{i<j}=\sum_{\langle i<j\rangle} + \sum_{\overline{\langle i<j\rangle}}$ is a constant term with respect to the update of $w$.
However, no known algorithm exists to solve a quadratic problem while maintaining isometric constraints.

\begin{figure*}[!t]
  \begin{center}
  \includegraphics[clip, width=0.8\textwidth]{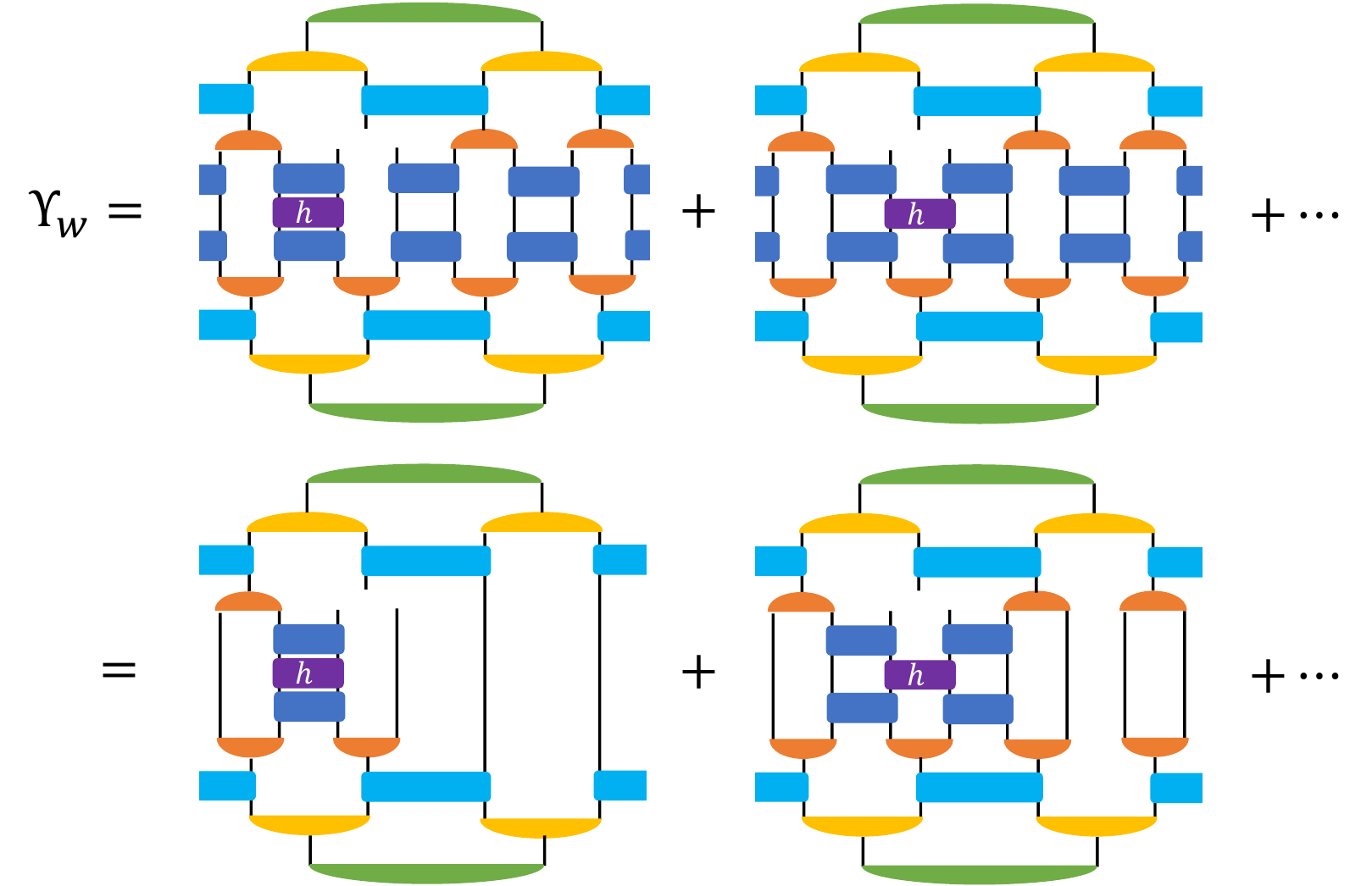}
  \caption{Diagram of the environment tensor $\Upsilon_w$ for tensor $w$. This is the summation of the tensor contraction, except $w$ from $e_{ij}$, and $(i, j)$ is involved with $w$ in a causal cone.} 
  \label{fig:environment_tensor}
  \end{center}
\end{figure*}

Evenbly and Vidal developed a method for optimizing a single tensor of MERA based on a linearizing approximation~\cite{Evenbly_Vidal_PhysRevB.79.144108}.
In this approach, we temporally regard $w$ and $w^{\dag}$ as independent tensors and optimize $w$ while keeping $w^{\dag}$ fixed, namely
\begin{eqnarray}
    \tilde{E}(w) & \equiv & \sum_{abc} \left[\Upsilon_w\right]^{c}_{ab} w^c_{ab} ~, \\ \left[\Upsilon_w\right]^{c'}_{a'b'} & \equiv & \sum_{abc}\sum_{\langle i<j \rangle} \left(w^{c}_{ab}\right)^* \left[\mathcal{E}_{ij}\right]^{abc}_{a'b'c'} ~,
\end{eqnarray}
where the tensor $\Upsilon_w$ is called the environment tensor of the $w$, for example, as shown in Fig~\ref{fig:environment_tensor}.

\begin{figure}[!h]
  \begin{center}
  \includegraphics[clip, width=3.0in]{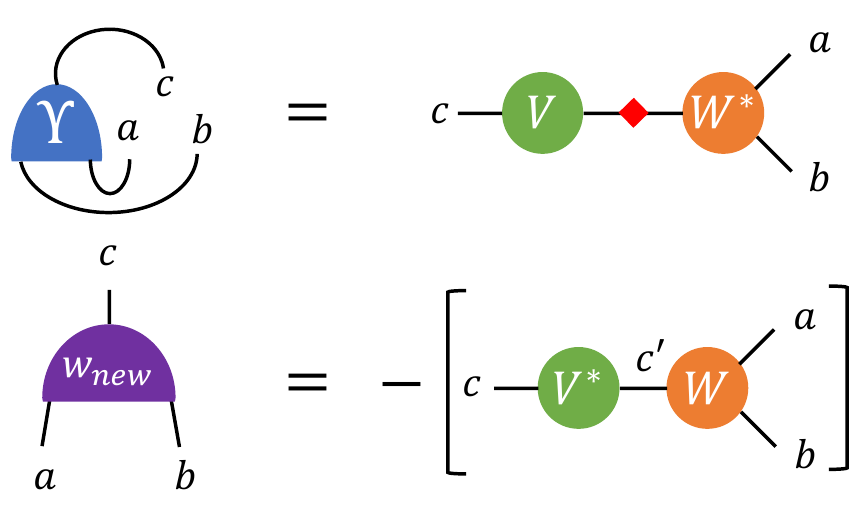}
  \caption{Schematic diagram of the update of $w$ with Evenbly-vidal algorithm equivalent to Eq.~(\ref{eq:update_w1}, \ref{eq:update_w2}). The red square tensor is the singler-value tensor $S$ in Eq.~(\ref{eq:update_w1}).}
  \label{fig:update_w}
  \end{center}
\end{figure}

To achieve a unique global minimization of $\tilde{E}(w)$, a singular value decomposition (SVD)
\begin{equation}\label{eq:update_w1}
\left[ \Upsilon_w \right]^{c}_{ab} \stackrel{\rm{SVD}}{=} \sum_{c'} V_{cc'}S_{c'}W^*_{c'ab}
\end{equation}
is applied to the environment tensor.
The tensor $w_{\text{new}}$ that minimizes $\tilde{E}(w_{\text{new}})$ is always given by
\begin{equation}\label{eq:update_w2}
\left[ w_{\text{new}} \right]^{c}_{ab} = -\sum_{c'} V^*_{cc'}W_{c'ab} ~,
\end{equation}
as like being refer to Fig.~\ref{fig:update_w}.
For this update to always work well, the Hamiltonian should be redefined as $\mathcal{H}_{\gamma} = \mathcal{H} - \gamma I$, where $\gamma$ is sufficiently large so that $\mathcal{H}_{\gamma}$ is negative definite.
However, the optimization step size is scaled by $\gamma^{-1}$~\cite{10.21468/SciPostPhys.10.2.040}, so we should choose $\gamma$ as small as possible practically.
In the Evenbly-Vidal method in non-uniform MERA, this linearized update is performed sequentially for each tensor that constructs the MERA.

Similarly, we can obtain $e_{ij}$ and optimize TN with the branching binary MERA; the typical numerical cost of evaluating $e_{ij}$ is reported to be $O(\chi^{12}N)$ if the system size is sufficiently large ~\cite{Evenbly_2014}.
Compared with the computational cost of MERA and branching MERA, the latter is exceedingly high, especially in higher-dimensional systems~\cite{PhysRevB.89.235113}. This has been the primary factor preventing us from analyzing the performance of branching MERA using classical computers.
However, in the case of quantum computers, there is no significant difference in the computational cost with respect to the number of gates. Specifically, the numbers of gates in the MERA and branching MERA are $O(N)$ and $O(N\log{N})$, respectively.

\subsection{Variational Quantum Eigensolver}\label{subsec:VQE}
The variational quantum eigensolver (VQE)~\cite{Peruzzo2014} is a quantum-classical hybrid variational method for finding the ground state energy of a target Hamiltonian $\mathcal{H}$ on an $N$-qubit system.
In the VQE, the variational wave function for the system is expressed as a product of parameterized quantum unitary gates
\begin{equation}
\ket{\Psi({\bm \theta})}=\prod_{k=1}^{K} \hat{u}_k(\bm{\theta}_k) \ket{0}^{\otimes N},
\end{equation}
where $\bm{\theta}=(\bm{\theta}_1,\bm{\theta}_2,\cdots,\bm{\theta}_K)$ and $\hat{u}_k(\bm{\theta}_k)$ means $k$-th quantum gate acting on single or multiple qubits specified by the user and has variational parameters $\bm{\theta}_k=(\theta_{k,1},\theta_{k,2},\cdots,\theta_{k,\ell_k})$; $\ell_k$ is the number of internal degrees of freedom for $\hat{u}_k$.
In this study, each $\hat{u}_k$ is an SU(4) gate, where $\bm{\theta}_k$ becomes a 15-dimensional real vector, acting on the $i_k$-th and $j_k$-th qubits with $1 \leq i_k < j_k \leq N$.
The variational wave function $\ket{\Psi}$ is always normalized due to the unitarity of the gates.

A goal of VQE is to minimize the variational energy
\begin{eqnarray}
E &=& \min_{\bm{\theta}} E(\bm{\theta}), \\
\textrm{where}~~ E(\bm{\theta}) &=& \ev{\mathcal{H}}{\Psi({\bm \theta})} \nonumber
\end{eqnarray}
with respect to $\bm{\theta}$. Since numerically exact values of the partial derivatives
\begin{equation} \label{eq:shift_rule}
\left\{ \partial_{\theta_{k,l}} E(\bm{\theta}) = \frac{ E(\bm{\theta} + \frac{\pi}{2} \bm{e}_{k,l}) - E(\bm{\theta} - \frac{\pi}{2} \bm{e}_{k,l}) }{2}  \right\}^{1\leq k \leq K}_{1\leq l \leq \ell_k}
\end{equation}
with the orthonormal unit vector $\bm{e}_{k,l}$ for shifting only the parameter $\theta_{k,l}$ are easily obtained by the parameter shift rule~\cite{Mitarai_PRA2018}, we can employ various gradient-based algorithms as optimizers of the VQE.
In practical NISQ applications, systematic perturbations linked with gate operations and statistical errors arising from the limited number of measurements influence energy evaluation.
However, we could take positive views that numerous methodologies for error mitigation have been developed to obtain the derivatives given by Eq.~(\ref{eq:shift_rule}).
In VQE, the calculation of expectation values $\{ E(\bm{\theta}), E(\bm{\theta} \pm \frac{\pi}{2} \bm{e}_{k,l}) \}$ is performed on a quantum computer because the calculation is a bottleneck in simulations using classical computers.
Subsequently, the classical computer updates the parameters using the gradient $\nabla E(\bm{\theta})$.
Of course, based on the parameter shift rules, we may choose an optimizer that does not use the gradient or the Hessian.
Finally, VQE achieves a ground-state search by iterative calculations between the procedures of classical and quantum computers.

The hyperparameters of the VQE are the initial values of the variational parameters, including the design of the quantum gate set, choice of the optimizer, and hyperparameters of the optimizer (see the review~\cite{TILLY20221} for more information).
These hyperparameters should be properly controlled according to the quantum computer's ability to perform high-precision calculations in VQE.
In particular, when using NISQ devices, it is essential to attempt to decrease the total number of quantum gates within an acceptable range of numerical accuracy to reduce the effect of noise.

\subsection{The synergistic framework}\label{subsec:synergistic_framework}
The problem with VQAs in NISQ devices is that statistical and systematic errors are associated with the measurement and noise, respectively.
A naive approach to solving this problem is to reduce the dependence on NISQ devices in existing quantum-classical hybrid algorithms.
For example, Okada et al. proposed an efficient protocol for the topological phase analysis of transverse field clusters and toric code models on quantum circuits~\cite{okada2022identification}.
In this protocol, all optimization of the quantum circuit, including the calculation of expectation values, is performed on a classical computer using causal cone structures, as discussed in Fig.~\ref{fig:causal_cone} for each local observable.
In addition, the expectation value evaluation of nonlocal observables, which are difficult to handle on a classical computer, is performed using the classically optimized quantum circuit.

However, this protocol has constraints on the target system and ansatz.
For example, the measured observables used in the optimization are local, and the ansatz comprises local gates with constant depth.
To avoid this difficulty, first, the synergistic framework described in Ref.~\cite{arxiv.2208.13673} employs the MPS after the variational calculation with TN methods on the classical computer and performs quantum circuit encoding of the MPS by sequential two-qubit-gate decomposition with 1D topology~\cite{rudolph2022decomposition, Ran_2020} as the initial ansatz of the VQE.
Second, the framework augments the quantum circuit encoded from the MPS with an additional parameterized quantum circuit.
In this step, the internal parameters of the additional quantum circuit are set to zero, so that they correspond to the identity operators.
Finally, the framework performs VQE with the augmented circuit as the initial ansatz.
The crucial properties of the framework can be summarized in the following two points: the VQE is trapped in the BP when all parameters of the augmented ansatz are given randomly but can avoid the BP by using the encoded quantum circuit; the all-to-all gate topology, which is different from the topology of the MPS, of the quantum circuit added after the quantum circuit encoding effectively contributes to better VQE performance.
These importance values are also reported in Ref. ~\cite{PhysRevX.12.011047, Grant2018}.

It is worth noting that in Ref.~\cite{arxiv.2208.13673} of the synergistic framework, the SU(4) gate product is created through a sequential process during the quantum-circuit encoding of the MPS.
This process makes it difficult to fully utilize classical computing power as a divide-and-conquer approach cannot be naively introduced.
Additionally, although all-to-all topology gates are most commonly used for entanglement augmentation, it is not a systematic procedure.
To achieve a quantum advantage in nonuniform systems, such as quantum chemical systems, by the synergistic framework, TN states can be easily converted to quantum circuits.
The entanglement structure of the variational quantum state must be changed by adding a few gates during the entanglement augmentation process.

\section{Entangled embedding VQE}
\label{sec:EEVQE}
To this end, we propose an entangled embedding VQE (EEVQE) protocol, using binary MERA for classical variational optimization and branching MERA augmented from binary MERA on a quantum circuit.
The protocol can be summarized in four steps:
\begin{enumerate}
\item Perform the variational optimization of the MERA state with the entropic area law on the classical computer for the target Hamiltonian. \label{item:classical_VQE}
\item Obtain the quantum circuit representation of the optimized MERA state using the technique of quantum circuit encoding. \label{item:QCE}
\item Embed the encoded circuit to an entanglement-augmented quantum circuit inspired by the branching MERA state with entropic volume law. \label{item:Embedding}
\item Perform VQE with the initial state given by the augmented circuit.\label{item:VQE}
\end{enumerate}
The following subsection will explain how to incorporate MERA states into a quantum circuit ansatz with the same degrees of freedom as the branching MERA state.
Additionally, we will introduce a modified version of the Evenbly-Vidal technique for updating MERA during classical variational optimization.
It will be used to compare optimizers for updating MERA in all-to-all coupled random systems, as discussed in Sec.~\ref{subsec:comparison_optimizer}.

\subsection{Quantum circuit representation of MERA and embedding process}\label{subsec:qc_rep}
\begin{figure}[!h]
    \begin{center}
    \includegraphics[clip, width=3.0in]{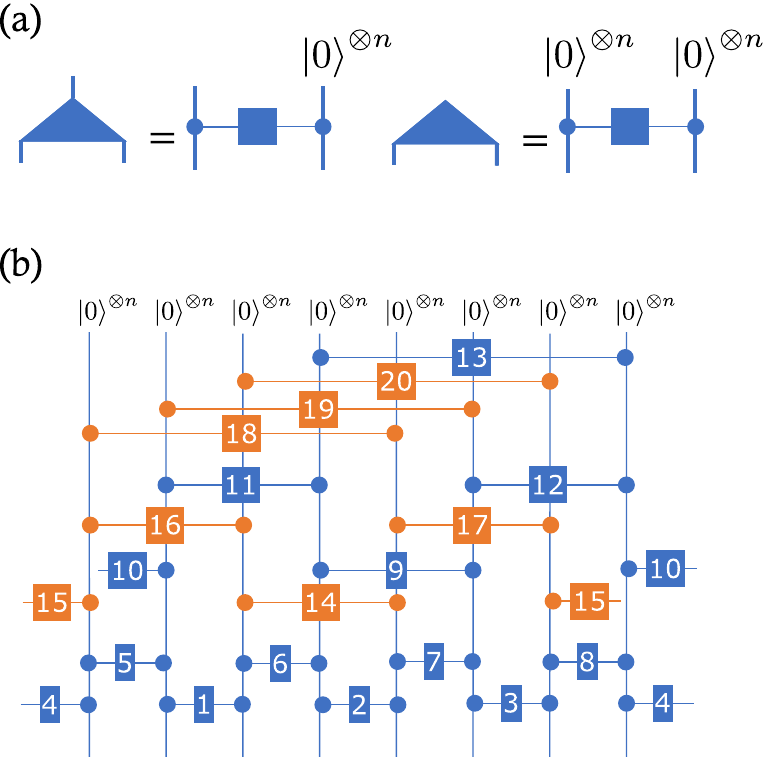}
    \caption{Schematic diagrams of (a) quantum circuit representations of disentangler, isometry, and top tensors, and (b) quantum circuit representation of the branching MERA for $N=8n$ with $\chi=2^n$ consisting of 20 SU($2^{2n}$) $2n$-qubit gates, where blue diagrams indicate the degree of freedom of (non-branching) MERA.}
    \label{fig:branchingMERA_quantum_circuit}
    \end{center}
\end{figure}

The disentangler $u$, isometry $w$, and top tensor $t$ with the bond dimension $\chi=2^n$ of MERA equivalent to the SU($2^{2n}$) rotation operator ($2n$-qubit gate) have different condition of input qubits $\ket{0}^{\otimes n}$ states as shown in Fig.~\ref{fig:branchingMERA_quantum_circuit}(a).
We should take a large $n$ in the case of complex problems and want more accuracy; for example, in \cite{Evenbly_Vidal_PhysRevB.79.144108} they use up to  $n=4$ roughly to benchmark for the translational symmetry 1D critical system, but the larger $n$ we take, the higher the computational cost for all EEVQE processes.
The quantum circuit representations of the binary MERA and branching MERA according to the above rule are depicted in Fig.~\ref{fig:branchingMERA_quantum_circuit}(b).
One of the processes of embedding the isometry and top tensors elements, which form the MERA state after variational optimization on a classical computer, into a unitary matrix of SU($2^{2n}$) involves using the modified Gram-Schmidt method (See Appendix.~\ref{sec:unitarization}).

In the case of $n=1$, by use of Cartan decomposition, any SU(4) operator $\hat{u}_{ij}$ for $i$- and $j$-th qubits can be decomposed into a kind of time evolution operator $\mathcal{D}_{ij}$ with XYZ-type interaction and SU(2) rotation operators $\mathcal{R}_{i}$ for one qubit~\cite{tucci2005introduction}, that means $u_{ij}=\mathcal{R}_i\mathcal{R}_j\mathcal{D}_{ij}\mathcal{R}'_i\mathcal{R}'_j$ where
\begin{equation}
\mathcal{D}_{ij}=e^{-{\rm i}{\bm k} \cdot {\bm \Sigma}_{ij}},~\mathcal{R}_i=e^{-{\rm i} \psi \sigma^z_i/2}e^{-{\rm i} \theta \sigma^y_i/2}e^{-{\rm i} \phi \sigma^z_i/2}
\end{equation}
with ${\bm k}=(k^x,k^y,k^z)$, ${\bm \Sigma}_{ij}=(\sigma_i^x\sigma_j^x,\sigma_i^y\sigma_j^y,\sigma_i^z\sigma_j^z)$; ${\bm \sigma}_i=(\sigma_i^x, \sigma_i^y, \sigma_i^z)$ are the Pauli operators for $i$-th qubit.

To execute $2n$-qubit gates equivalent to the unitary matrix on a quantum computer, decomposing the $2n$-qubit gates into a product of two-qubit SU(4) gates is essential because once we have the product, we can break down each SU(4) gate into CNOT gates and single-qubit rotational gates~\cite{PhysRevA.63.062309}, which are commonly used in NISQ today.
Computer-assisted search and numerical optimization for the circuit decomposition have been widely studied~\cite{DiVincenzo_PHYCMP1994,Amy_TCAD2013,Nam_npjqi2018,Khatri_Q2019,Nagarajan_QCS2021,younis2021qfast,fosel2021quantum,AQCE,meister2022exploring,rakyta2022efficient,nemkov2023efficient, Smith_2023}.
We can encode the MERA state in a quantum circuit by using decomposition techniques on each SU($2^{2n}$) gate individually and simultaneously in classical parallel computing.

To explain the embedding process in detail, we refer to Fig.~\ref{fig:branchingMERA_quantum_circuit}(b), where the MERA state is embedded through quantum gates within the blue gate. The branching MERA is represented by a quantum circuit with an additional orange quantum gate alongside the blue gate. The embedding process is accomplished by setting the internal parameters such that all orange circuits function as identity operators. In the case of the VQE calculation, with the circuit as the initial condition, getting stuck in a local solution, we introduce weak noise to the orange (and blue) quantum gates to avoid such a situation~\cite{arxiv.2208.13673,PhysRevX.12.011047}.

Studying the effectiveness of branching MERA through a quantum computer is beneficial, as it entails lower computational costs than MERA simulations on a classical computer.
MERA requires an order of SU($2^{2n}$) gates of $O(N)$ while branching MERA requires $O(N \log N)$.
These costs refer to the computational requirements for producing a quantum state in a quantum computer by operating quantum gates on individual qubits.
However, suppose that the computer can compute mutually commutative quantum gates that are spatially separated. Both computational costs can be reduced to $O(\log N)$, which corresponds to the number of layers of the isometry and the disentangler.
Consequently, the relative computational cost of branching MERA is similar to that of MERA.

\subsection{Improved Procedures for Sequential Tensor Optimization in MERA}\label{sec:improved procedures}
Since this study mainly focuses on applications to all-to-all coupled random systems, the classical computational part of the optimization of the inhomogeneous MERA becomes important and nontrivial.
We modify the optimization method (Evenbly-Vidal method) employed in the original paper~\cite{Evenbly_Vidal_PhysRevB.79.144108} to handle this situation.
This procedure updates the tensor to be optimized using the singular vector obtained by singular value decomposition of the environment tensor constructed by all tensor contractions except the tensor to be optimized.
However, we must perform a considerable number of sweep optimizations and may be trapped in a local minimum with this method, even for MERA networks with sufficient bond dimensions $\chi$ to represent the ground state of the target system, reflecting the effect of linearizing the tensor optimization.
Therefore, it is unfavorable for EEVQE.

An approach to reduce the number would be to perform diagonalization of the effective Hamiltonian without linearizing the cost function with respect to only the top tensor of the MERA network.
The effective Hamiltonian is defined as follows:
Suppose that a quantum circuit $C$ providing the MERA state $\ket{\Psi}=C \ket{0}^{\otimes N}$ is shown in Fig.~\ref{fig:branchingMERA_quantum_circuit}(b) with $n=1$, and we consider updating the top tensor $u_{13}$ labeled as No. 13.
Then, we introduce $C'=Cu^{\dagger}_{13}$ and $\Psi'_{mn} = \bra{m,n} u_{13}^{} \ket{0}^{\otimes 2}$ with $m,n \in \{0,1\}$ and rewrite $\ket{\Psi}$ and the expectation value $E=\bra{\Psi} \mathcal{H} \ket{\Psi}$ as follows
\begin{eqnarray}
\ket{\Psi} & = & \sum_{m,n} \Psi'_{mn} C' \ket{0}^{\otimes N-2} \ket{m,n}~, \\
E & = & \sum_{m,n,m',n'} \Psi_{mn}^{\prime *} \mathcal{H}'_{mn,m'n'} \Psi'_{m'n'}~,
\end{eqnarray}
where
\begin{equation}
\mathcal{H}'_{mn,m'n'} = \bra{m,n} \bra{0}^{\otimes N-2} C^{\prime \dagger} \mathcal{H} C' \ket{0}^{\otimes N-2} \ket{m',n'}~
\end{equation}
is the effective Hamiltonian referred to when updating the internal degrees of freedom of $u_{13}$.

\begin{figure}
  \begin{center}
    \includegraphics[width = 80mm]{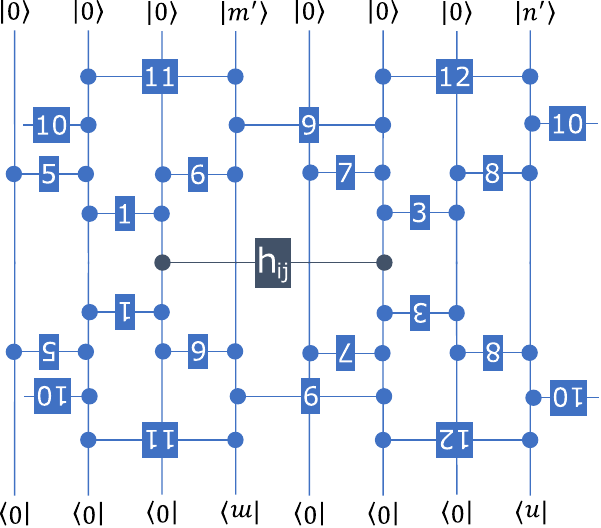}
    \caption{Diagrammatic representation of an effective local two-body Hamiltonian $h^{\prime (ij)}_{mn,m'n'}$ for MERA as in Fig.~\ref{fig:branchingMERA_quantum_circuit}(b) with $(i,j)=(3,6)$ and $\chi=2$. Quantum circuits labeled with numbers 2 and 4 are omitted because they do not contribute to the effective local Hamiltonian.}
    \label{fig:effective_hamiltonian}
    \end{center}
\end{figure}

Here, we assume that the total Hamiltonian $\mathcal{H}=\sum_{i<j} h_{ij}$ is given by the sum of the two-body Hamiltonian $h_{ij}$ between $i$- and $j$-th sites.
Then, we can evaluate the effective Hamiltonian $\mathcal{H}'$ required to optimize circuit 13 as the sum of the local two-body effective Hamiltonian
\begin{equation}
\left( h^{\prime}_{ij} \right)_{mn,m'n'}=\bra{m,n} \bra{0}^{\otimes N-2} C^{\prime \dagger} h_{ij} C' \ket{0}^{\otimes N-2} \ket{m',n'},
\end{equation}
and its diagrammatic representation
is shown in Fig.~\ref{fig:effective_hamiltonian}.
After the contractions, we perform the diagonalization of the effective Hamiltonian $\mathcal{H}'$ and the eigenvectors $\{\Phi'_{mn}\}$ corresponding to the ground states of $\mathcal{H}'$ are computed.
Finally, the update of circuit 13 is completed by embedding
\begin{equation}
   \left( u_{13} \right)^{0,0}_{mn}:=\Phi'_{mn}
\end{equation}
and performing the unitarization as noted in Sec.~\ref{subsec:qc_rep}.

It should be mentioned here that such embedding of eigenvectors obtained by the diagonalization of the effective Hamiltonian into the tensor to be optimized is difficult to perform for tensors other than the top tensor because the MERA network, unlike the tree tensor network (TTN)~\cite{PhysRevA.74.022320,PhysRevB.80.235127,PhysRevB.82.205105}, contains a loop structure.
On the other hand, diagonalization applies to optimizing parameterized quantum gates that act only on the zero kets of the initial state, regardless of the details of the variational quantum circuit ansatz.

\section{Numerical simulations}
\label{sec:numerical_sim}

\subsection{Hamiltonians and numerical setup}
In order to investigate the properties of our procedure, we consider the following all-to-all coupled random transverse field Ising model
\begin{equation}
    H_{\textrm{Ising}} = \sum_{i<j}{J_{ij}\sigma^z_i\sigma^z_j} + \sum_{i}{h_{i}\sigma^x_i}~, \label{eq:ham_Ising}
\end{equation}
XYZ model
\begin{equation}
    H_{\textrm{XYZ}} = \sum_{i<j} {\bm J}_{ij} \cdot {\bm \Sigma}_{ij},
\label{eq:ham_XYZ}
\end{equation}
and Heisenberg model in a random field
\begin{equation}
    H_{\textrm{Heisenberg}} = \sum_{i<j} J_{ij} \bm 1 \cdot {\bm \Sigma}_{ij}
    + \sum_{i} {\bm h}_i \cdot {\bm \sigma}_i
\label{eq:ham_HB}
\end{equation}
with ${\bm 1}=(1,1,1)$, where the coupling constants $J_{ij}$, ${\bm J}_{ij}=(J^x_{ij},J^y_{ij},J^z_{ij})$ and the magnetic fields $h_i$ and ${\bm h}_i=(h^x_i,h^y_i,h^z_i)$ are given by uniform random numbers in the range [-1,1).
In the actual calculation, the Hamiltonians are pre-processed to be negative definite by shifting the constant terms $\gamma_{ij}$, which is defined as $\gamma_{ij}=\max[f_{ij},0]$ with $f_{ij}$ as the maximum eigenvalue of $h_{ij}$, as discussed in Sec.~\ref{subsec:tn_opt}.

The quantity evaluated hereafter is the relative error from the exact ground state energy $E_{\rm exact}$
\begin{equation}
\Delta = (E - E_{\rm exact})/E_{\rm exact}
\end{equation}
and its random-averaged values $\bar{\Delta}$, where we prepare $10^2$ realizations for the random coupling constants and magnetic fields.
Then, the initial parameters of the MERA state for each realization are common random values.

For the variational update of the MERA state on the classical computer, we adopt the modified Evenbly-Vedal method as discussed in Sec.~\ref{sec:improved procedures} to benefit from the TN method on the classical computer (See Sec.~\ref{subsec:comparison_optimizer}).
The sequential update schedule for the tensors was set from the bottom to the top layer, as shown in Fig.~\ref{fig:binaryMERA, isometric_condition}(a), adopting a left-to-right order within each layer, and the number of iterations for the sequential update is up to $10^3$.
In this study, we use the Julia version of the ITensor library~\cite{itensor} for TN calculations and only consider $\chi=2$ to focus on the exact circuit encoding of the MERA state.

Then, we use the quantum-simulation software Cirq~\cite{cirq} for the quantum circuit encoding of the SU(4) unitary matrix using Cartan's decomposition.
Note that in this study, the $\chi = 2$ tensors are analytically encoded exactly into 2-qubit gates so that we can parallelize the procedure regardless of the target structure.
The VQE procedure employs the BFGS method, which is a quasi-Newtonian method.
The hyperparameter is only the number of BFGS iterations in this step, and the number is taken up to $10^4$ in our study.

In this study, we focus on how well the VQE calculation of branching MERA with the optimized MERA  as the initial wave function is performed when the effect of noise is not considered, so we utilize a quantum circuit simulator qulacs~\cite{Suzuki2021qulacsfast} and the numerical analysis software library scipy~\cite{2020SciPy-NMeth} for the VQE calculations.

\subsection{Benchmarks of MERA optimization procedure}\label{subsec:comparison_optimizer}
\begin{figure}[!h]
    \centering
    \includegraphics[clip, width=3.4in]{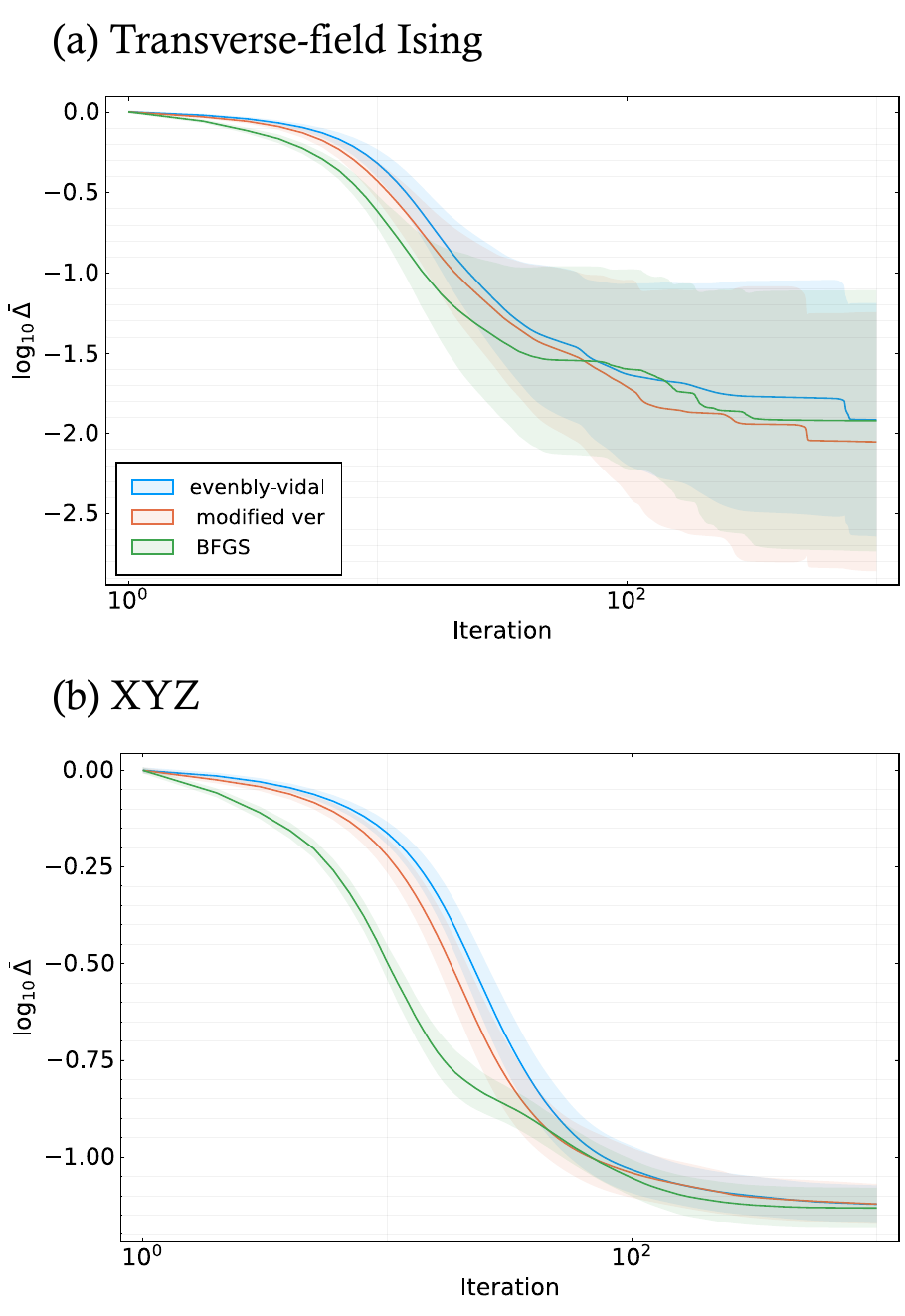}
    \caption{
    Iteration dependence of $\bar{\Delta}$ for all-to-all coupled random (a) Ising and (b) XYZ systems with $(\chi,N)=(2,16)$.
    }
    \label{fig:benchmark about optimizer}
\end{figure}

Since classical optimization of inhomogeneous MERA is important, as mentioned before, we first compare which optimization method performs better in three optimization methods - Evenbly-Vidal, modified Evenbly-Vidal, and BFGS in the random Ising/XYZ model with all-to-all couplings.
The results are shown in Fig.~\ref{fig:benchmark about optimizer}, where the colored lines and the light-shaded region represent $\bar{\Delta}$ and its variance, respectively.

The results of our study demonstrate that the modified Evenbly-Vidal method exhibits a marginally superior convergence speed compared to the original method in both models.
In addition, our findings indicate that the modified method enhances the convergence error in the transverse Ising model.
Moreover, we compared the modified Evenbly-Vidal method and the BFGS for the variational optimization of the MERA state.
The results show that BFGS is more effective in reducing the variational energy during the initial optimization phase than the modified Evenbly-Vidal method for all models analyzed.
This agrees with the expectation that the sequential optimization based on the TN method may be relatively easy to trap in local minima.

Also, we briefly confirm that the performance of the modified Evenbly-Vidal method improves as the weight of the degree of freedom of the top tensors in TN states increases in Appendix.~\ref{sec:diagonalization}.

\subsection{EEVQE calculations}
\begin{figure*}[!p]
  \centering
  \includegraphics[clip, width=6.0in]{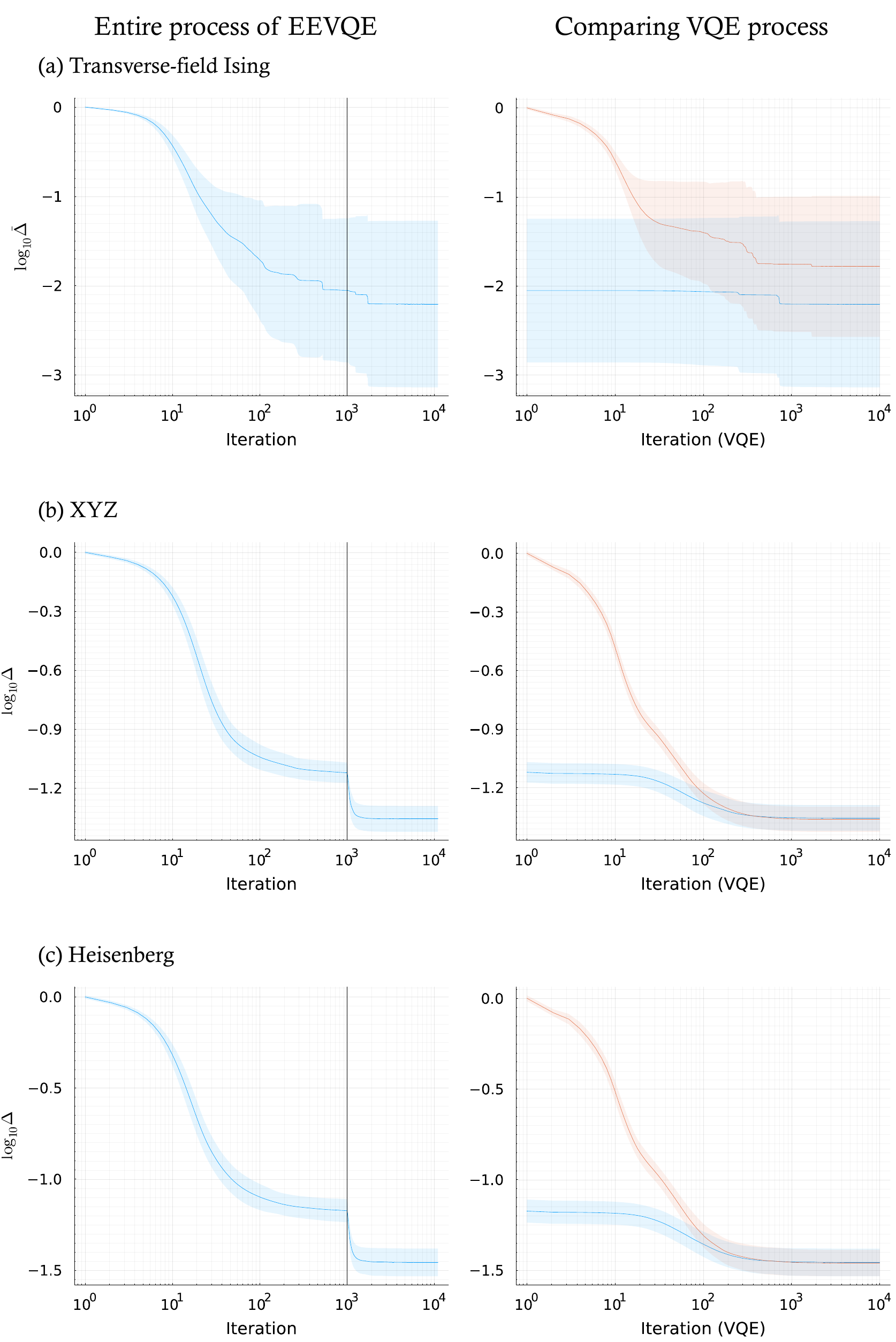}
  \caption{
    Iteration dependence of the random-averaged relative error $\bar{\Delta}$ in the all-to-all coupled random (a) transverse-field Ising, (b) XYZ, and (c) Heisenberg models (left panels), and comparison of $\bar{\Delta}$ in the VQE calculation using initial states with embedded MERA states after TN optimizations and random branching MERA states (right panels).
  }
  \label{fig:Iteration vs Error}
\end{figure*}

In this subsection, we conduct benchmark EEVQE calculations by augmenting gates to form the branching MERA and executing the VQE for the ground-state search of Hamiltonians in Eqs.~(\ref{eq:ham_Ising})-(\ref{eq:ham_HB}).
The left panels of Fig.~\ref{fig:Iteration vs Error} show the iteration dependence of $\bar{\Delta}$ obtained by the proposed procedure for each Hamiltonian.
The vertical black line at $10^3$ iterations represents the switching from the variational optimization of the MERA state to the VQE calculation with the branching-MERA ansatz.

In all three models, the optimization of the MERA state is almost completed up to $10^3$ iterations; in fact, we confirm that this claim is valid by increasing the number of iterations up to $5 \times 10^3$.
After the quantum circuit encoding of the MERA state, we perform the VQE calculation, and the $\log_{10} \bar{\Delta}$ decreases by 0.15, 0.23, and 0.28 in the all-to-all coupled random Ising model, XYZ model, and Heisenberg model reflecting the entanglement augmentations.
In particular, since the XYZ and Heisenberg models have more quantum fluctuation, the intensity of off-diagonal components may be comparable with its diagonal components, and the accuracy improvements by the entanglement augmentation are more significant than for the Ising model.


The right panel of Fig.~\ref{fig:Iteration vs Error} focuses on the VQE calculation and compares the behaviors of $\bar{\Delta}$ with two initial conditions: optimized MERA states and random branching MERA states.
In the case of the all-to-all coupled random transverse-field Ising model, there is a moderately high initial-state dependence on the converged $\bar{\Delta}$, and the VQE with a random branching MERA as the initial state is trapped by local minima.
It is a specific example where the EEVQE results show an advantage over the standard VQE.
On the other hand, in the case of the all-to-all coupled random XYZ and Heisenberg models, there is no significant difference in the converged $\bar{\Delta}$ for both initial conditions.
However, up to approximately $10^2$ iterations of VQE, EEVQE can reduce $\bar{\Delta}$ compared to VQE, where the random branching MERA states are the initial states.
This result suggests that the short-time VQE computation using the NISQ device is superior in our procedure, which is consistent with the current trend in quantum algorithm development.

\section{Conclusion and Discussion}\label{sec:summary}
We have developed an entangled embedded VQE (EEVQE) method that uses a branching binary MERA with a one-dimensional entropic volume law as the ansatz for VQE calculations.
In the EEVQE method, the binary MERA is optimized using a modified Evenbly-Vidal method on a classical computer and serves as the initial state for VQE.
Unlike the original synergistic frameworks~\cite{arxiv.2208.13673} that use the TN structure only for the initial state, EEVQE incorporates an entanglement augmentation topology based on the TN structure during gate addition.
We investigated the performance of our method on the all-to-all coupled random transverse field Ising, XYZ, and Heisenberg models, and evaluated the random average $\bar{\Delta}$ of the relative error between the variational energy and the exact ground state energy under computational conditions $(\chi, N)=(2, 16)$ to obtain the exact quantum circuit representation of the branching MERA embedded with the optimized MERA and the exact ground state of each model.

In the numerical simulation, we first examined the benchmark of the MERA optimization procedure on the classical computer using three methods: the Evenbly-Vidal algorithm, its modification, and the BFGS method as a fundamental knowledge of MERA for non-uniform systems.
The results show that the BFGS method is superior to the other methods for all the models studied.
This trend is considered a phenomenon specific to non-uniform systems since similar analysis in uniform systems~\cite{PhysRevX.12.011047} shows that sequential local updating by the Evenbly-Vidal method is superior to the gradient-based methods.
This property becomes crucial when using the synergistic framework to study the ground states of quantum chemical calculations, which is one of the goals of VQE for all-to-all coupled inhomogeneous systems.

Second, we verified that the EEVQE method reduces $\log_{10} \bar{\Delta}$ by 0.15, 0.23, 0.28 in the all-to-all coupled random Ising, XYZ, and Heisenberg models, reﬂecting the entanglement augmentations while being free from the initialization problem of VQE calculations.
Furthermore, we confirm that EEVQE can prevent becoming stuck in a local minimum in the all-to-all coupled random Ising model. At the same time, the VQE with a randomly initialized branching-MERA ansatz may become trapped in the minimum.
Also, a comparison of VQE with random initialized branching MERA and branching MERA with embedded an optimized binary MERA as the initial state shows that the latter is superior to the former up to about $10^2$ iterations of the update in VQE.
In Appendix~\ref{seq:rainbow_state}, we also experimented with the EEVQE for the state that yields the entropic volume law to reinforce the validity of the previous results.
These properties meet the requirements of the quantum variational algorithm using the NISQ device, which is to improve the accuracy of the best solution on a classical computer by a small amount of use of a quantum computer.

In the present study, since we applied one-dimensional branching MERA to a non-uniform system with all-to-all coupled interactions, obtaining a highly accurate $\bar{\Delta}$ was not easy with $\chi=2$, even for the TN that satisfies the entropic volume law.
It is not because of the barren plateau but because one-dimensional MERA and branching MERA with $\chi = 2$ cannot represent the target ground state.
Even branching MERA has a log-scale depth of the circuit, so it is considered not to be trapped in the barren plateau.
Therefore, there remains to consider a scheme that can handle a $\chi > 2$ in order to find out the limits.
In this case, it is possible to apply automatic quantum circuit encoding (AQCE)~\cite{AQCE} or other circuit decomposition techniques ~\cite{DiVincenzo_PHYCMP1994,Amy_TCAD2013,Nam_npjqi2018,Khatri_Q2019,Nagarajan_QCS2021,younis2021qfast,fosel2021quantum,AQCE,meister2022exploring,rakyta2022efficient,nemkov2023efficient, Smith_2023} for each disentangler, isometry, and top tensor of the TN to perform divide-and-conquer quantum circuit encoding for the entire TN.

As another future research, we can perform structural optimization of TN for each random interaction.
In this case, $\bar{\Delta}$ with higher accuracy is expected to be easily obtained even for TN with the current minimum degree of freedom, $\chi=2$.
Furthermore, we aim to integrate this approach into quantum circuits constructed using the aforementioned encoding methodologies to enhance the precision of it.
We anticipate that these schemes can be applied to sets of quantum gates representing each isometric tensor individually, rather than to entire circuits.
There have been reports on the structure optimization for TTNs that do not include loop structures in the TN~\cite{Li_2022,HikiharaPRR2023,okunishi2023ptep}.
However, they cannot be extend directly to MERA, including loop structures, and further research is needed.

In addition, a deep MERA (DMERA)~\cite{DMERA} has been reported as an algorithm that mixes the roles of the disentangler and isometry in MERA and improves the performance of MERA through quantum circuit representation.
We can also introduce the branching degree of freedom in the DMERA.
By examining the TN microscopically in terms of the product of two-qubit gates, it is more necessary to design a TN/quantum circuit structure that matches the geometrical aspect of the entanglement contained in the target state.

Other methods not treated in this study as optimizers for updating MERA include learning rate~\cite{rudolph2022decomposition}, automatic differentiation~\cite{Geng_2022}, and Riemannian optimization~\cite{10.21468/SciPostPhys.10.2.040}. It is interesting to verify whether these methods perform better than the BFGS method in the all-to-all coupled non-uniform models as one of the future issues to be addressed.

Finally, we are considering the development of quantum computers and their applications ~\cite{Cong2019,PRXQuantum.2.010342}, and the verification of EEVQE on actual NiSQ devices should be addressed in future studies.

\begin{acknowledgments}
This work is partially supported by
KAKENHI Grant Numbers JP22H01171, JP21H04446,
and a Grant-in-Aid for Transformative Research Areas "The Natural Laws of Extreme
Universe---A New Paradigm for Spacetime and Matter from Quantum Information"
(KAKENHI Grant Nos. JP21H05182, JP21H05191) from JSPS of Japan.
It is also supported by JST PRESTO No. JPMJPR1911, MEXT Q-LEAP Grant No.
JPMXS0120319794, and JST COI-NEXT No. JPMJPF2014.
H.U was supported by the COE research grant in computational science from Hyogo
Prefecture and Kobe City through Foundation for Computational Science.
We are grateful for allocating computational resources of the HOKUSAI BigWaterfall supercomputing system at RIKEN and SQUID at the Cybermedia Center, Osaka University.
\end{acknowledgments}

\appendix

\section{Unitarization of isometry by Modified Gram-Schmidt method}\label{sec:unitarization}
We consider the isometry $w=\{w^c_{ab}\}_{1\leq a,b \leq \chi}^{1\leq c \leq \chi'}$, where $\chi'$ satisfies $1 \leq \chi' \leq \chi^2$, with a given bond dimension $\chi$ of a positive integer under the isometric condition in Eq.~(\ref{eq:isometric_w}).
For unitarization of the isometry $w$, we introduce a four-rank tensor $u=\{u^{cd}_{ab}\}_{1\leq a,b \leq \chi}^{1 \leq c \leq \chi', 1 \leq d \leq \chi^2/\chi'}$ and embed $w$ and a random tensor $r=\{r^{cm}_{ab}\}_{1\leq a,b \leq \chi}^{1 \leq c \leq \chi', 2 \leq m \leq \chi^2/\chi'}$ into $u$ as
\begin{equation}
    u^{cd}_{ab}=\left\{
    \begin{matrix}
        w^c_{ab} & (d=1) \\
        r^{cd}_{ab} & (2\leq d \leq \chi^2/\chi') \\
    \end{matrix}
    \right. ~.
\end{equation}
Then, we perform the modified Gram-Schmidt procedure in Algorithm \ref{alg:MGS} and obtain the isometric tensor $u$ that satisfies the isometric conditions in Eq.~(\ref{eq:isometric_u_1}) and Eq.~(\ref{eq:isometric_u_2}).
\begin{figure}[!h]
\begin{algorithm}[H]
    \caption{the modified Gram-Schmidt procedures}
    \label{alg:MGS}
    \begin{algorithmic}[1]
    \FOR{$c=1:\chi'$,~$d=2:\chi^2/\chi'$}
        \FOR{$n=1:c+\chi'(d-1)-1$}
            \STATE find $c'$ and $d'$ that satisfy $n=c'+\chi'(d'-1)$
            \STATE $u_{:,:}^{cd}:=u_{:,:}^{cd}- \left(\sum_{a,b}  u_{ab}^{c'd'*} u_{ab}^{cd} \right) u_{:,:}^{c'd'}$
            \STATE (Note that $u_{:,:}^{cd}=\{u_{ab}^{cd}\}_{1\leq a,b \leq \chi }$.)
        \ENDFOR
        \STATE $u_{:,:}^{cd}:=u_{:,:}^{cd}/|| u_{:,:}^{cd} ||_2$
    \ENDFOR
    \end{algorithmic}
\end{algorithm}
\end{figure}

\section{Network and the bond dimension dependence of the performance of modified Evenbly-Vidal method}\label{sec:diagonalization}
\begin{figure*}
  \centering
    \includegraphics[clip, width=6in]{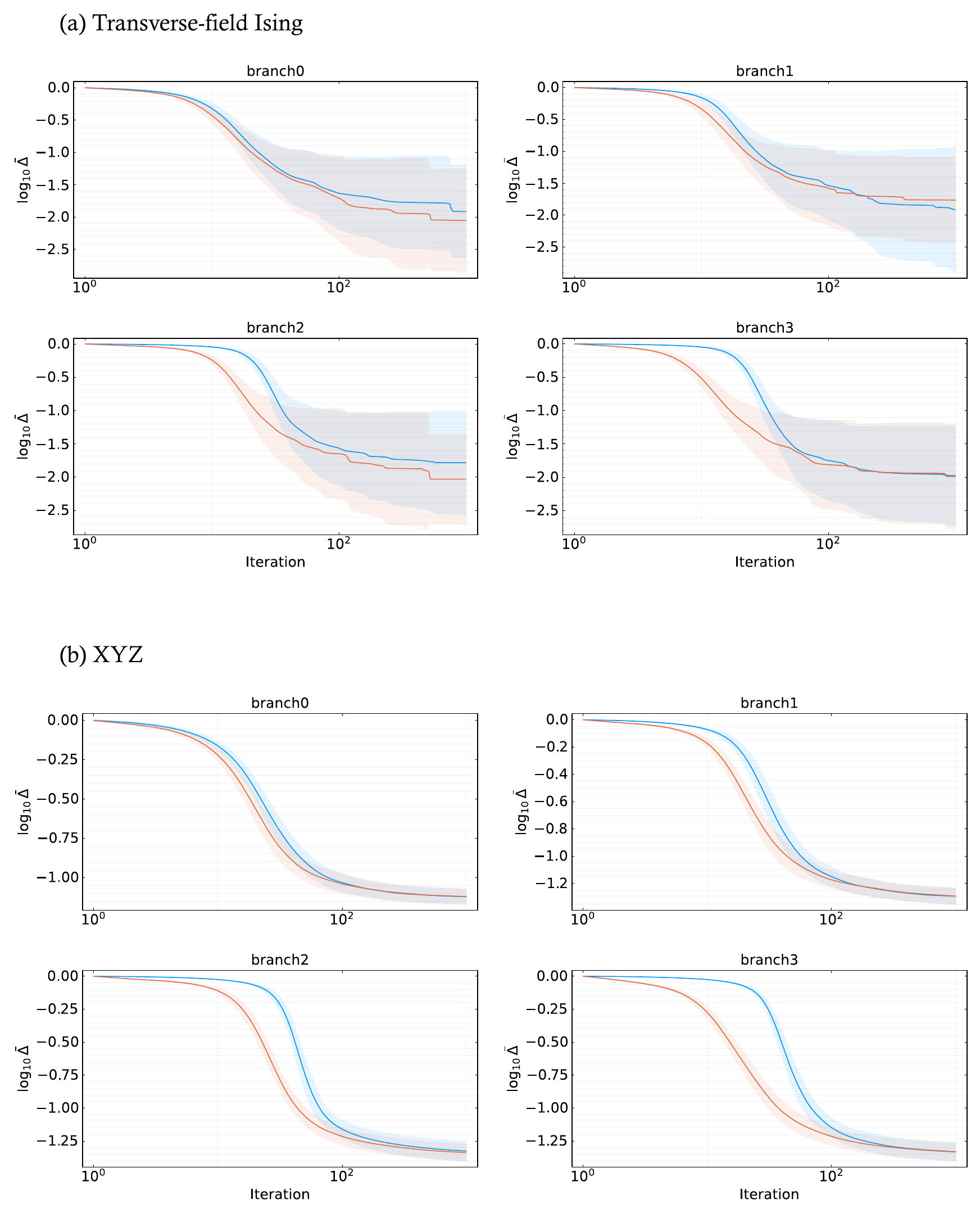}
  \caption{
  Iteration dependence of $\bar{\Delta}$ for all-to-all coupled random (a) Ising and (b) XYZ systems with $(\chi,N)=(2,16)$. The blue and orange lines represent the case of the Evenbly-Vidal method and its modification, as introduced in Sec.~\ref{sec:improved procedures}, respectively. The branching pattern was named according to the procedure shown in Fig.~\ref{fig:branching}.
  }
  \label{fig:diagonalization}
\end{figure*}

In Fig.~\ref{fig:diagonalization}, we compared the performance of the original Evenbly-Vidal method and modified one for inhomogeneous MERAs with all branching patterns, as shown in Fig.~\ref{fig:branching}.
Our results indicate that, in all model and branching pattern combinations, the modified method outperforms the original Evenbly-Vidal method during the initial optimization iterations up to $10^2$.
Additionally, this improvement is more significant in both models as the number of top tensors increases due to bifurcations.

\begin{figure*}
  \centering
    \includegraphics[clip, width=6in]{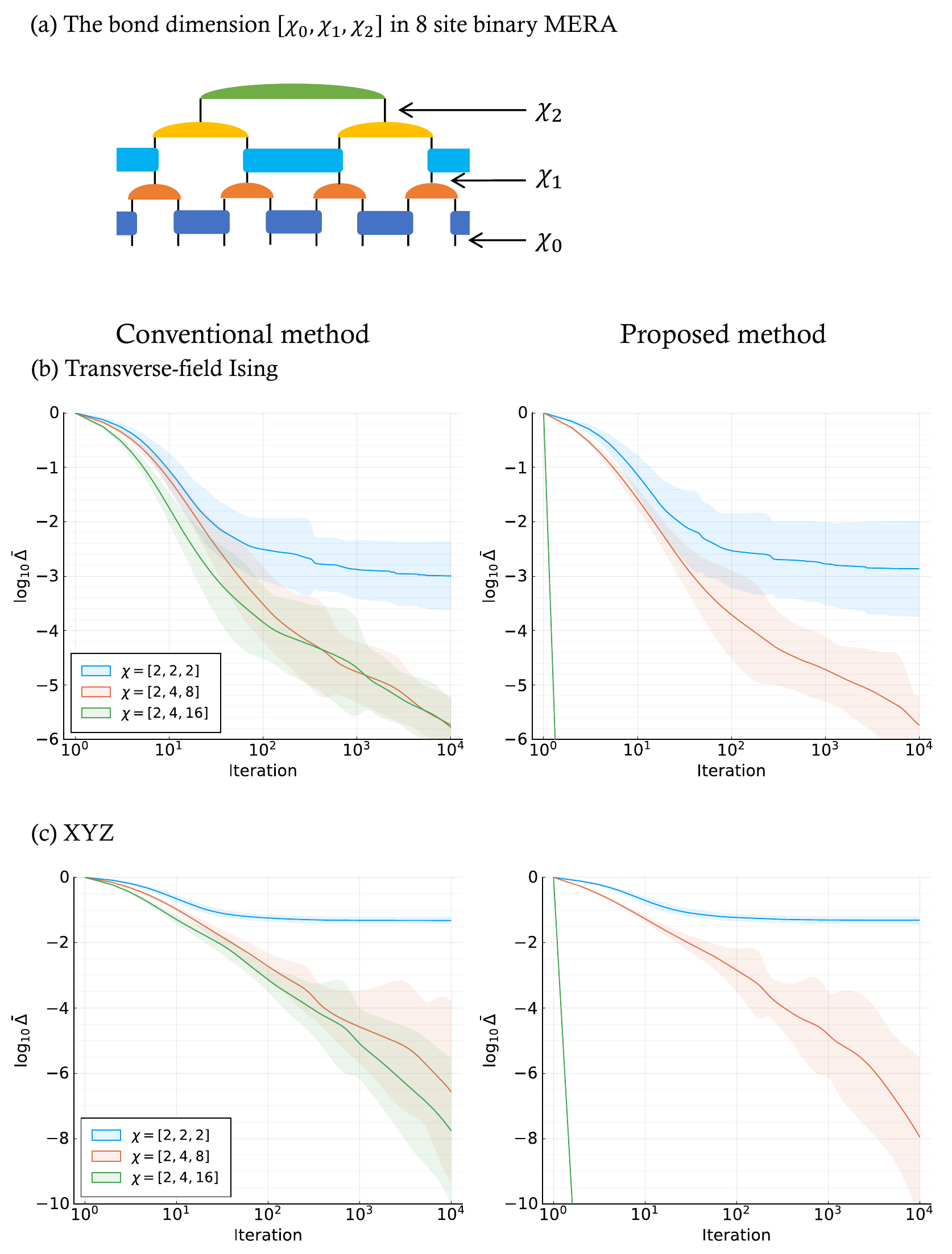}
  \caption{The legend $\chi = \left[ \circ, \circ, \circ \right]$ in Fig~\ref{fig:diagonalization2}~(b) and \ref{fig:diagonalization2}~(c) means the $\chi = \left[ \chi_0, \chi_1, \chi_2 \right]$ as shown in Fig.~\ref{fig:diagonalization2}~(a). The $\chi_0$ is fixed as $2$ in case of the spin system.
  The results displayed in the left and right panels correspond to the conventional method and our proposed method, respectively.}
  \label{fig:diagonalization2}
\end{figure*}

Also in Fig.~\ref{fig:diagonalization2}, we conducted a comparison of the bond dimension dependencies between the two methods in $8$-site system.
The results show that, by increasing the bond dimension, the modified method is superior to the original method.
Particularly, in bond dimensions $\chi = \left[ 2, 4, 16 \right]$, which is max bond dimension in $8$ size system, means it covers whole hilbert space, using the modified method we can get the exact solution at one time, but with the original method the convergence speeds are almost same as in $\chi = \left[ 2, 4, 8 \right]$
This result suggests that the modified method is more effective in the case of large bond dimensions.

\section{Result of EEVQE for the rainbow state}\label{seq:rainbow_state}
\begin{figure}
  \centering
  \includegraphics[clip, width=3.4in]{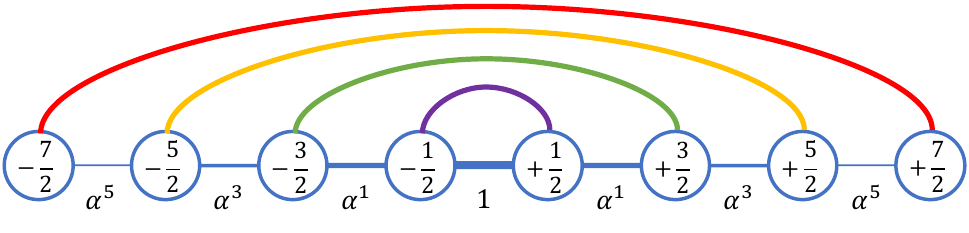}
  \caption{
  Representation of the rainbow state showing $(-l, l)$ valence bonds above the central link.
  The nearest neighbor interactions are given in terms of $\alpha=\exp(-h)$.
  }
  \label{fig:rainbow_state}
\end{figure}

In this section, we apply EEVQE to a 1D inhomogeneous Heisenberg chain whose Hamiltonian is defined as
\begin{equation}
\hat{H} = \bm{\sigma}_{-\frac{1}{2}} \cdot \bm{\sigma}_{\frac{1}{2}} + \sum_{l=\frac{1}{2}}^{(N-3)/2} e^{-2hl} \left[ \bm{\sigma}_{l} \cdot \bm{\sigma}_{l+1} + \bm{\sigma}_{-l} \cdot \bm{\sigma}_{-(l+1)} \right] ~,
  \label{eq:rainbow-heisenberg}
\end{equation}
where $h \geq 0$ denotes an inhomogeneity parameter.
For $h = 0$, the system is nothing but the uniform Heisenberg chain with open boundary conditions.
In the strong inhomogeneity limit $h \gg 1$, the ground state of $\hat{H}$ becomes the rainbow state~\cite{Vitagliano_2010, Ramirez_2014, Ramirez_2015}, which is a valence bond solid composed of singlet bonds connecting opposite sites of the chain, as shown in Fig.~\ref{fig:rainbow_state}.
The rainbow state has an entropic volume law with respect to dividing the system into two parts, left and right, in one dimension.
In this system, we can expect that the improvement in the numerical accuracy of the entanglement augmentation by EEVQE becomes significant with respect to $h$.

\begin{figure*}
  \centering
  \includegraphics[clip, width=5.25in]{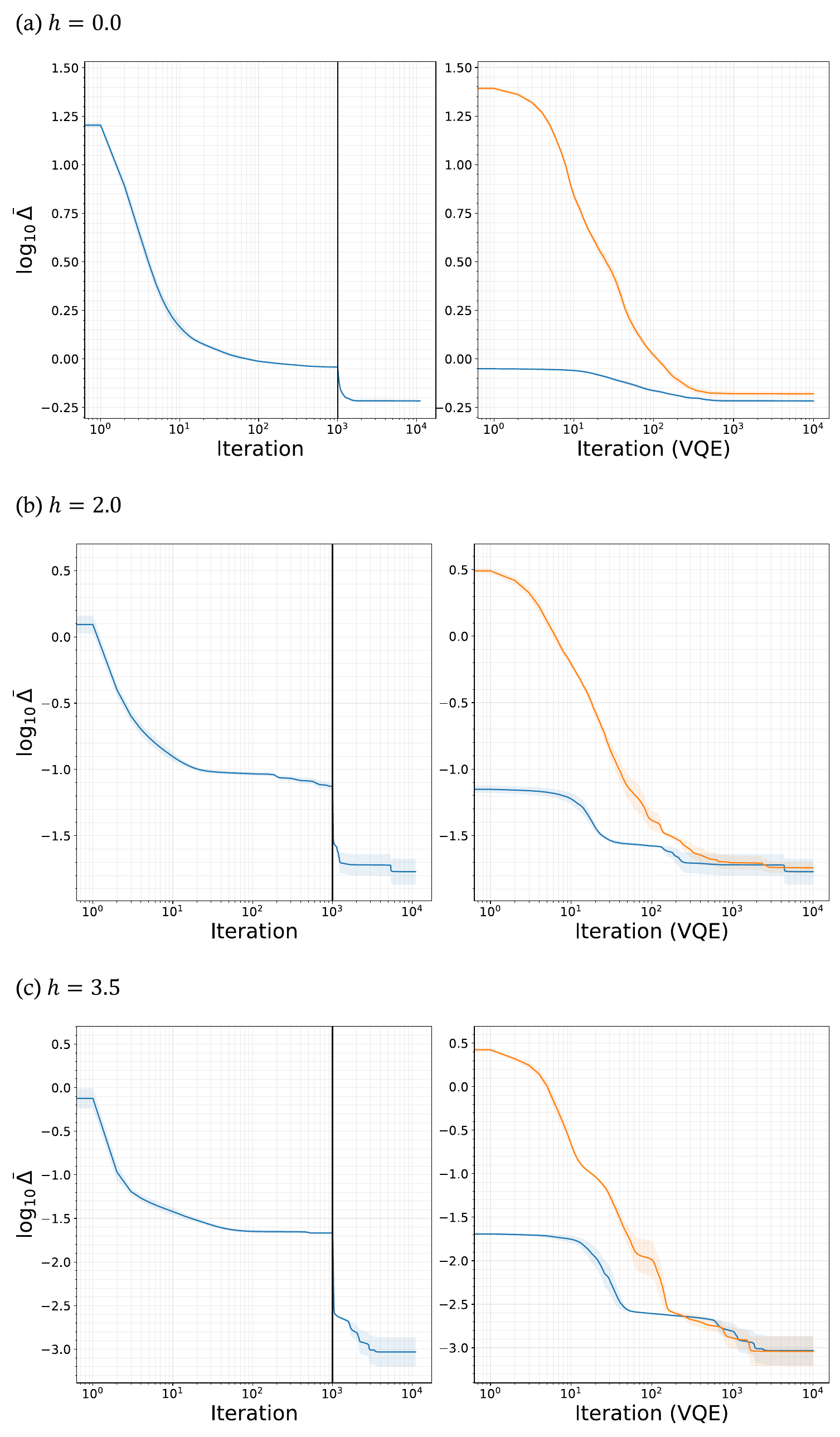}
  \caption{
    The result of the EEVQE solution for the inhomogeneous Heisenberg model as \ref{eq:rainbow-heisenberg} with $N=16$.
  }
  \label{fig:eevqe_rainbow}
\end{figure*}

The results of EEVQE for $h=0.0, 2.0, 3.5$ are shown in the left panels of Fig.~\ref{fig:eevqe_rainbow}.
As we expected, the difference between $\log_{10}\bar{\Delta}$ for the MERA with $10^3$ iterations and that for the branching MERA with $10^4$ iterations increases monotonically with respect to $h$, and $\log_{10}\bar{\Delta}$ for the branching MERA with $10^4$ iterations decreasing monotonically with respect to $h$. 
As in the random XYZ/Heisenberg model case in Sec.~\ref{sec:numerical_sim}, we confirm that the EEVQE solution outperformed the random initial VQE solution up to approximately $10^{2}$ iterations in the right panels of Fig. \ref{fig:eevqe_rainbow}.

\bibliography{main}

\begin{thebibliography}{64}%
\makeatletter
\providecommand \@ifxundefined [1]{%
 \@ifx{#1\undefined}
}%
\providecommand \@ifnum [1]{%
 \ifnum #1\expandafter \@firstoftwo
 \else \expandafter \@secondoftwo
 \fi
}%
\providecommand \@ifx [1]{%
 \ifx #1\expandafter \@firstoftwo
 \else \expandafter \@secondoftwo
 \fi
}%
\providecommand \natexlab [1]{#1}%
\providecommand \enquote  [1]{``#1''}%
\providecommand \bibnamefont  [1]{#1}%
\providecommand \bibfnamefont [1]{#1}%
\providecommand \citenamefont [1]{#1}%
\providecommand \href@noop [0]{\@secondoftwo}%
\providecommand \href [0]{\begingroup \@sanitize@url \@href}%
\providecommand \@href[1]{\@@startlink{#1}\@@href}%
\providecommand \@@href[1]{\endgroup#1\@@endlink}%
\providecommand \@sanitize@url [0]{\catcode `\\12\catcode `\$12\catcode `\&12\catcode `\#12\catcode `\^12\catcode `\_12\catcode `\%12\relax}%
\providecommand \@@startlink[1]{}%
\providecommand \@@endlink[0]{}%
\providecommand \url  [0]{\begingroup\@sanitize@url \@url }%
\providecommand \@url [1]{\endgroup\@href {#1}{\urlprefix }}%
\providecommand \urlprefix  [0]{URL }%
\providecommand \Eprint [0]{\href }%
\providecommand \doibase [0]{https://doi.org/}%
\providecommand \selectlanguage [0]{\@gobble}%
\providecommand \bibinfo  [0]{\@secondoftwo}%
\providecommand \bibfield  [0]{\@secondoftwo}%
\providecommand \translation [1]{[#1]}%
\providecommand \BibitemOpen [0]{}%
\providecommand \bibitemStop [0]{}%
\providecommand \bibitemNoStop [0]{.\EOS\space}%
\providecommand \EOS [0]{\spacefactor3000\relax}%
\providecommand \BibitemShut  [1]{\csname bibitem#1\endcsname}%
\let\auto@bib@innerbib\@empty
\bibitem [{\citenamefont {Daskin}\ and\ \citenamefont {Kais}(2018)}]{Daskin_2018}%
  \BibitemOpen
  \bibfield  {author} {\bibinfo {author} {\bibfnamefont {A.}~\bibnamefont {Daskin}}\ and\ \bibinfo {author} {\bibfnamefont {S.}~\bibnamefont {Kais}},\ }\bibfield  {title} {\bibinfo {title} {Direct application of the phase estimation algorithm to find the eigenvalues of the hamiltonians},\ }\href {https://doi.org/10.1016/j.chemphys.2018.01.002} {\bibfield  {journal} {\bibinfo  {journal} {Chem. Phys.}\ }\textbf {\bibinfo {volume} {514}},\ \bibinfo {pages} {87} (\bibinfo {year} {2018})}\BibitemShut {NoStop}%
\bibitem [{\citenamefont {Low}\ and\ \citenamefont {Chuang}(2019)}]{Low_2019}%
  \BibitemOpen
  \bibfield  {author} {\bibinfo {author} {\bibfnamefont {G.~H.}\ \bibnamefont {Low}}\ and\ \bibinfo {author} {\bibfnamefont {I.~L.}\ \bibnamefont {Chuang}},\ }\bibfield  {title} {\bibinfo {title} {Hamiltonian simulation by qubitization},\ }\href {https://doi.org/10.22331/q-2019-07-12-163} {\bibfield  {journal} {\bibinfo  {journal} {Quantum}\ }\textbf {\bibinfo {volume} {3}},\ \bibinfo {pages} {163} (\bibinfo {year} {2019})}\BibitemShut {NoStop}%
\bibitem [{\citenamefont {Gily{\'{e}}n}\ \emph {et~al.}(2019)\citenamefont {Gily{\'{e}}n}, \citenamefont {Su}, \citenamefont {Low},\ and\ \citenamefont {Wiebe}}]{Gily_n_2019}%
  \BibitemOpen
  \bibfield  {author} {\bibinfo {author} {\bibfnamefont {A.}~\bibnamefont {Gily{\'{e}}n}}, \bibinfo {author} {\bibfnamefont {Y.}~\bibnamefont {Su}}, \bibinfo {author} {\bibfnamefont {G.~H.}\ \bibnamefont {Low}},\ and\ \bibinfo {author} {\bibfnamefont {N.}~\bibnamefont {Wiebe}},\ }\bibfield  {title} {\bibinfo {title} {Quantum singular value transformation and beyond: exponential improvements for quantum matrix arithmetics},\ }in\ \href {https://doi.org/10.1145/3313276.3316366} {\emph {\bibinfo {booktitle} {Proceedings of the 51st Annual {ACM} {SIGACT} Symposium on Theory of Computing}}}\ (\bibinfo  {publisher} {{ACM}},\ \bibinfo {year} {2019})\ p.\ \bibinfo {pages} {193–204}\BibitemShut {NoStop}%
\bibitem [{\citenamefont {Preskill}(2018)}]{Preskill2018quantumcomputingin}%
  \BibitemOpen
  \bibfield  {author} {\bibinfo {author} {\bibfnamefont {J.}~\bibnamefont {Preskill}},\ }\bibfield  {title} {\bibinfo {title} {Quantum {C}omputing in the {NISQ} era and beyond},\ }\href {https://doi.org/10.22331/q-2018-08-06-79} {\bibfield  {journal} {\bibinfo  {journal} {{Quantum}}\ }\textbf {\bibinfo {volume} {2}},\ \bibinfo {pages} {79} (\bibinfo {year} {2018})}\BibitemShut {NoStop}%
\bibitem [{\citenamefont {Cerezo}\ \emph {et~al.}(2021)\citenamefont {Cerezo}, \citenamefont {Arrasmith}, \citenamefont {Babbush}, \citenamefont {Benjamin}, \citenamefont {Endo}, \citenamefont {Fujii}, \citenamefont {McClean}, \citenamefont {Mitarai}, \citenamefont {Yuan}, \citenamefont {Cincio},\ and\ \citenamefont {Coles}}]{Cerezo2021}%
  \BibitemOpen
  \bibfield  {author} {\bibinfo {author} {\bibfnamefont {M.}~\bibnamefont {Cerezo}}, \bibinfo {author} {\bibfnamefont {A.}~\bibnamefont {Arrasmith}}, \bibinfo {author} {\bibfnamefont {R.}~\bibnamefont {Babbush}}, \bibinfo {author} {\bibfnamefont {S.~C.}\ \bibnamefont {Benjamin}}, \bibinfo {author} {\bibfnamefont {S.}~\bibnamefont {Endo}}, \bibinfo {author} {\bibfnamefont {K.}~\bibnamefont {Fujii}}, \bibinfo {author} {\bibfnamefont {J.~R.}\ \bibnamefont {McClean}}, \bibinfo {author} {\bibfnamefont {K.}~\bibnamefont {Mitarai}}, \bibinfo {author} {\bibfnamefont {X.}~\bibnamefont {Yuan}}, \bibinfo {author} {\bibfnamefont {L.}~\bibnamefont {Cincio}},\ and\ \bibinfo {author} {\bibfnamefont {P.~J.}\ \bibnamefont {Coles}},\ }\bibfield  {title} {\bibinfo {title} {Variational quantum algorithms},\ }\href {https://doi.org/10.1038/s42254-021-00348-9} {\bibfield  {journal} {\bibinfo  {journal} {Nat. Rev. Phys.}\ }\textbf {\bibinfo {volume} {3}},\ \bibinfo {pages} {625} (\bibinfo {year} {2021})}\BibitemShut {NoStop}%
\bibitem [{\citenamefont {Peruzzo}\ \emph {et~al.}(2014)\citenamefont {Peruzzo}, \citenamefont {McClean}, \citenamefont {Shadbolt}, \citenamefont {Yung}, \citenamefont {Zhou}, \citenamefont {Love}, \citenamefont {Aspuru-Guzik},\ and\ \citenamefont {O'Brien}}]{Peruzzo2014}%
  \BibitemOpen
  \bibfield  {author} {\bibinfo {author} {\bibfnamefont {A.}~\bibnamefont {Peruzzo}}, \bibinfo {author} {\bibfnamefont {J.}~\bibnamefont {McClean}}, \bibinfo {author} {\bibfnamefont {P.}~\bibnamefont {Shadbolt}}, \bibinfo {author} {\bibfnamefont {M.-H.}\ \bibnamefont {Yung}}, \bibinfo {author} {\bibfnamefont {X.-Q.}\ \bibnamefont {Zhou}}, \bibinfo {author} {\bibfnamefont {P.~J.}\ \bibnamefont {Love}}, \bibinfo {author} {\bibfnamefont {A.}~\bibnamefont {Aspuru-Guzik}},\ and\ \bibinfo {author} {\bibfnamefont {J.~L.}\ \bibnamefont {O'Brien}},\ }\bibfield  {title} {\bibinfo {title} {A variational eigenvalue solver on a photonic quantum processor},\ }\href {https://doi.org/10.1038/ncomms5213} {\bibfield  {journal} {\bibinfo  {journal} {Nat. Commun.}\ }\textbf {\bibinfo {volume} {5}},\ \bibinfo {pages} {4213} (\bibinfo {year} {2014})}\BibitemShut {NoStop}%
\bibitem [{\citenamefont {McClean}\ \emph {et~al.}(2018)\citenamefont {McClean}, \citenamefont {Boixo}, \citenamefont {Smelyanskiy}, \citenamefont {Babbush},\ and\ \citenamefont {Neven}}]{McClean2018}%
  \BibitemOpen
  \bibfield  {author} {\bibinfo {author} {\bibfnamefont {J.~R.}\ \bibnamefont {McClean}}, \bibinfo {author} {\bibfnamefont {S.}~\bibnamefont {Boixo}}, \bibinfo {author} {\bibfnamefont {V.~N.}\ \bibnamefont {Smelyanskiy}}, \bibinfo {author} {\bibfnamefont {R.}~\bibnamefont {Babbush}},\ and\ \bibinfo {author} {\bibfnamefont {H.}~\bibnamefont {Neven}},\ }\bibfield  {title} {\bibinfo {title} {Barren plateaus in quantum neural network training landscapes},\ }\href {https://doi.org/10.1038/s41467-018-07090-4} {\bibfield  {journal} {\bibinfo  {journal} {Nat. Commun.}\ }\textbf {\bibinfo {volume} {9}},\ \bibinfo {pages} {4812} (\bibinfo {year} {2018})}\BibitemShut {NoStop}%
\bibitem [{\citenamefont {Grant}\ \emph {et~al.}(2019)\citenamefont {Grant}, \citenamefont {Wossnig}, \citenamefont {Ostaszewski},\ and\ \citenamefont {Benedetti}}]{Grant2019initialization}%
  \BibitemOpen
  \bibfield  {author} {\bibinfo {author} {\bibfnamefont {E.}~\bibnamefont {Grant}}, \bibinfo {author} {\bibfnamefont {L.}~\bibnamefont {Wossnig}}, \bibinfo {author} {\bibfnamefont {M.}~\bibnamefont {Ostaszewski}},\ and\ \bibinfo {author} {\bibfnamefont {M.}~\bibnamefont {Benedetti}},\ }\bibfield  {title} {\bibinfo {title} {An initialization strategy for addressing barren plateaus in parametrized quantum circuits},\ }\href {https://doi.org/10.22331/q-2019-12-09-214} {\bibfield  {journal} {\bibinfo  {journal} {{Quantum}}\ }\textbf {\bibinfo {volume} {3}},\ \bibinfo {pages} {214} (\bibinfo {year} {2019})}\BibitemShut {NoStop}%
\bibitem [{\citenamefont {Okada}\ \emph {et~al.}(2022)\citenamefont {Okada}, \citenamefont {Osaki}, \citenamefont {Mitarai},\ and\ \citenamefont {Fujii}}]{okada2022identification}%
  \BibitemOpen
  \bibfield  {author} {\bibinfo {author} {\bibfnamefont {K.~N.}\ \bibnamefont {Okada}}, \bibinfo {author} {\bibfnamefont {K.}~\bibnamefont {Osaki}}, \bibinfo {author} {\bibfnamefont {K.}~\bibnamefont {Mitarai}},\ and\ \bibinfo {author} {\bibfnamefont {K.}~\bibnamefont {Fujii}},\ }\href@noop {} {\bibinfo {title} {Identification of topological phases using classically-optimized variational quantum eigensolver}} (\bibinfo {year} {2022}),\ \Eprint {https://arxiv.org/abs/2202.02909} {arXiv:2202.02909 [quant-ph]} \BibitemShut {NoStop}%
\bibitem [{\citenamefont {Rudolph}\ \emph {et~al.}(2022{\natexlab{a}})\citenamefont {Rudolph}, \citenamefont {Miller}, \citenamefont {Chen}, \citenamefont {Acharya},\ and\ \citenamefont {Perdomo-Ortiz}}]{arxiv.2208.13673}%
  \BibitemOpen
  \bibfield  {author} {\bibinfo {author} {\bibfnamefont {M.~S.}\ \bibnamefont {Rudolph}}, \bibinfo {author} {\bibfnamefont {J.}~\bibnamefont {Miller}}, \bibinfo {author} {\bibfnamefont {J.}~\bibnamefont {Chen}}, \bibinfo {author} {\bibfnamefont {A.}~\bibnamefont {Acharya}},\ and\ \bibinfo {author} {\bibfnamefont {A.}~\bibnamefont {Perdomo-Ortiz}},\ }\href@noop {} {\bibinfo {title} {Synergy between quantum circuits and tensor networks: Short-cutting the race to practical quantum advantage}} (\bibinfo {year} {2022}{\natexlab{a}}),\ \Eprint {https://arxiv.org/abs/2208.13673} {arXiv:2208.13673 [quant-ph]} \BibitemShut {NoStop}%
\bibitem [{\citenamefont {Orús}(2014)}]{ORUS2014117}%
  \BibitemOpen
  \bibfield  {author} {\bibinfo {author} {\bibfnamefont {R.}~\bibnamefont {Orús}},\ }\bibfield  {title} {\bibinfo {title} {A practical introduction to tensor networks: Matrix product states and projected entangled pair states},\ }\href {https://doi.org/https://doi.org/10.1016/j.aop.2014.06.013} {\bibfield  {journal} {\bibinfo  {journal} {Ann. Phys.}\ }\textbf {\bibinfo {volume} {349}},\ \bibinfo {pages} {117} (\bibinfo {year} {2014})}\BibitemShut {NoStop}%
\bibitem [{\citenamefont {Or{\'u}s}(2014)}]{Orus2014}%
  \BibitemOpen
  \bibfield  {author} {\bibinfo {author} {\bibfnamefont {R.}~\bibnamefont {Or{\'u}s}},\ }\bibfield  {title} {\bibinfo {title} {Advances on tensor network theory: symmetries, fermions, entanglement, and holography},\ }\href {https://doi.org/10.1140/epjb/e2014-50502-9} {\bibfield  {journal} {\bibinfo  {journal} {Eur. Phys. J. B}\ }\textbf {\bibinfo {volume} {87}},\ \bibinfo {pages} {280} (\bibinfo {year} {2014})}\BibitemShut {NoStop}%
\bibitem [{\citenamefont {Ran}\ \emph {et~al.}(2020)\citenamefont {Ran}, \citenamefont {Tirrito}, \citenamefont {Peng}, \citenamefont {Chen}, \citenamefont {Tagliacozzo}, \citenamefont {Su},\ and\ \citenamefont {Lewenstein}}]{ran2020tensor}%
  \BibitemOpen
  \bibfield  {author} {\bibinfo {author} {\bibfnamefont {S.-J.}\ \bibnamefont {Ran}}, \bibinfo {author} {\bibfnamefont {E.}~\bibnamefont {Tirrito}}, \bibinfo {author} {\bibfnamefont {C.}~\bibnamefont {Peng}}, \bibinfo {author} {\bibfnamefont {X.}~\bibnamefont {Chen}}, \bibinfo {author} {\bibfnamefont {L.}~\bibnamefont {Tagliacozzo}}, \bibinfo {author} {\bibfnamefont {G.}~\bibnamefont {Su}},\ and\ \bibinfo {author} {\bibfnamefont {M.}~\bibnamefont {Lewenstein}},\ }\href@noop {} {\emph {\bibinfo {title} {Tensor network contractions: methods and applications to quantum many-body systems}}}\ (\bibinfo  {publisher} {Springer Nature},\ \bibinfo {year} {2020})\BibitemShut {NoStop}%
\bibitem [{\citenamefont {Okunishi}\ \emph {et~al.}(2022)\citenamefont {Okunishi}, \citenamefont {Nishino},\ and\ \citenamefont {Ueda}}]{ONU2022TN}%
  \BibitemOpen
  \bibfield  {author} {\bibinfo {author} {\bibfnamefont {K.}~\bibnamefont {Okunishi}}, \bibinfo {author} {\bibfnamefont {T.}~\bibnamefont {Nishino}},\ and\ \bibinfo {author} {\bibfnamefont {H.}~\bibnamefont {Ueda}},\ }\bibfield  {title} {\bibinfo {title} {Developments in the tensor network — from statistical mechanics to quantum entanglement},\ }\href {https://doi.org/10.7566/JPSJ.91.062001} {\bibfield  {journal} {\bibinfo  {journal} {J. Phys. Soc. Jpn.}\ }\textbf {\bibinfo {volume} {91}},\ \bibinfo {pages} {062001} (\bibinfo {year} {2022})}\BibitemShut {NoStop}%
\bibitem [{\citenamefont {Liu}\ \emph {et~al.}(2019)\citenamefont {Liu}, \citenamefont {Zhang}, \citenamefont {Wan},\ and\ \citenamefont {Wang}}]{Liu_PRR2019}%
  \BibitemOpen
  \bibfield  {author} {\bibinfo {author} {\bibfnamefont {J.-G.}\ \bibnamefont {Liu}}, \bibinfo {author} {\bibfnamefont {Y.-H.}\ \bibnamefont {Zhang}}, \bibinfo {author} {\bibfnamefont {Y.}~\bibnamefont {Wan}},\ and\ \bibinfo {author} {\bibfnamefont {L.}~\bibnamefont {Wang}},\ }\bibfield  {title} {\bibinfo {title} {Variational quantum eigensolver with fewer qubits},\ }\href {https://doi.org/10.1103/PhysRevResearch.1.023025} {\bibfield  {journal} {\bibinfo  {journal} {Phys. Rev. Res.}\ }\textbf {\bibinfo {volume} {1}},\ \bibinfo {pages} {023025} (\bibinfo {year} {2019})}\BibitemShut {NoStop}%
\bibitem [{\citenamefont {Haghshenas}\ \emph {et~al.}(2022)\citenamefont {Haghshenas}, \citenamefont {Gray}, \citenamefont {Potter},\ and\ \citenamefont {Chan}}]{PhysRevX.12.011047}%
  \BibitemOpen
  \bibfield  {author} {\bibinfo {author} {\bibfnamefont {R.}~\bibnamefont {Haghshenas}}, \bibinfo {author} {\bibfnamefont {J.}~\bibnamefont {Gray}}, \bibinfo {author} {\bibfnamefont {A.~C.}\ \bibnamefont {Potter}},\ and\ \bibinfo {author} {\bibfnamefont {G.~K.-L.}\ \bibnamefont {Chan}},\ }\bibfield  {title} {\bibinfo {title} {Variational power of quantum circuit tensor networks},\ }\href {https://doi.org/10.1103/PhysRevX.12.011047} {\bibfield  {journal} {\bibinfo  {journal} {Phys. Rev. X}\ }\textbf {\bibinfo {volume} {12}},\ \bibinfo {pages} {011047} (\bibinfo {year} {2022})}\BibitemShut {NoStop}%
\bibitem [{\citenamefont {Miao}\ and\ \citenamefont {Barthel}(2023)}]{miao2023convergence}%
  \BibitemOpen
  \bibfield  {author} {\bibinfo {author} {\bibfnamefont {Q.}~\bibnamefont {Miao}}\ and\ \bibinfo {author} {\bibfnamefont {T.}~\bibnamefont {Barthel}},\ }\href@noop {} {\bibinfo {title} {Convergence and quantum advantage of trotterized mera for strongly-correlated systems}} (\bibinfo {year} {2023}),\ \Eprint {https://arxiv.org/abs/2303.08910} {arXiv:2303.08910 [quant-ph]} \BibitemShut {NoStop}%
\bibitem [{\citenamefont {White}(1992)}]{PhysRevLett.69.2863}%
  \BibitemOpen
  \bibfield  {author} {\bibinfo {author} {\bibfnamefont {S.~R.}\ \bibnamefont {White}},\ }\bibfield  {title} {\bibinfo {title} {Density matrix formulation for quantum renormalization groups},\ }\href {https://doi.org/10.1103/PhysRevLett.69.2863} {\bibfield  {journal} {\bibinfo  {journal} {Phys. Rev. Lett.}\ }\textbf {\bibinfo {volume} {69}},\ \bibinfo {pages} {2863} (\bibinfo {year} {1992})}\BibitemShut {NoStop}%
\bibitem [{\citenamefont {Schollwöck}(2011)}]{SCHOLLWOCK201196}%
  \BibitemOpen
  \bibfield  {author} {\bibinfo {author} {\bibfnamefont {U.}~\bibnamefont {Schollwöck}},\ }\bibfield  {title} {\bibinfo {title} {The density-matrix renormalization group in the age of matrix product states},\ }\href {https://doi.org/https://doi.org/10.1016/j.aop.2010.09.012} {\bibfield  {journal} {\bibinfo  {journal} {Ann. Phys.}\ }\textbf {\bibinfo {volume} {326}},\ \bibinfo {pages} {96} (\bibinfo {year} {2011})},\ \bibinfo {note} {january 2011 Special Issue}\BibitemShut {NoStop}%
\bibitem [{\citenamefont {Zauner-Stauber}\ \emph {et~al.}(2018)\citenamefont {Zauner-Stauber}, \citenamefont {Vanderstraeten}, \citenamefont {Fishman}, \citenamefont {Verstraete},\ and\ \citenamefont {Haegeman}}]{PhysRevB.97.045145}%
  \BibitemOpen
  \bibfield  {author} {\bibinfo {author} {\bibfnamefont {V.}~\bibnamefont {Zauner-Stauber}}, \bibinfo {author} {\bibfnamefont {L.}~\bibnamefont {Vanderstraeten}}, \bibinfo {author} {\bibfnamefont {M.~T.}\ \bibnamefont {Fishman}}, \bibinfo {author} {\bibfnamefont {F.}~\bibnamefont {Verstraete}},\ and\ \bibinfo {author} {\bibfnamefont {J.}~\bibnamefont {Haegeman}},\ }\bibfield  {title} {\bibinfo {title} {Variational optimization algorithms for uniform matrix product states},\ }\href {https://doi.org/10.1103/PhysRevB.97.045145} {\bibfield  {journal} {\bibinfo  {journal} {Phys. Rev. B}\ }\textbf {\bibinfo {volume} {97}},\ \bibinfo {pages} {045145} (\bibinfo {year} {2018})}\BibitemShut {NoStop}%
\bibitem [{\citenamefont {Eisert}\ \emph {et~al.}(2010)\citenamefont {Eisert}, \citenamefont {Cramer},\ and\ \citenamefont {Plenio}}]{RevModPhys.82.277}%
  \BibitemOpen
  \bibfield  {author} {\bibinfo {author} {\bibfnamefont {J.}~\bibnamefont {Eisert}}, \bibinfo {author} {\bibfnamefont {M.}~\bibnamefont {Cramer}},\ and\ \bibinfo {author} {\bibfnamefont {M.~B.}\ \bibnamefont {Plenio}},\ }\bibfield  {title} {\bibinfo {title} {Colloquium: Area laws for the entanglement entropy},\ }\href {https://doi.org/10.1103/RevModPhys.82.277} {\bibfield  {journal} {\bibinfo  {journal} {Rev. Mod. Phys.}\ }\textbf {\bibinfo {volume} {82}},\ \bibinfo {pages} {277} (\bibinfo {year} {2010})}\BibitemShut {NoStop}%
\bibitem [{\citenamefont {Goldsborough}\ and\ \citenamefont {Evenbly}(2017)}]{Goldsborough_PRB2017}%
  \BibitemOpen
  \bibfield  {author} {\bibinfo {author} {\bibfnamefont {A.~M.}\ \bibnamefont {Goldsborough}}\ and\ \bibinfo {author} {\bibfnamefont {G.}~\bibnamefont {Evenbly}},\ }\bibfield  {title} {\bibinfo {title} {Entanglement renormalization for disordered systems},\ }\href {https://doi.org/10.1103/PhysRevB.96.155136} {\bibfield  {journal} {\bibinfo  {journal} {Phys. Rev. B}\ }\textbf {\bibinfo {volume} {96}},\ \bibinfo {pages} {155136} (\bibinfo {year} {2017})}\BibitemShut {NoStop}%
\bibitem [{\citenamefont {Grant}\ \emph {et~al.}(2018)\citenamefont {Grant}, \citenamefont {Benedetti}, \citenamefont {Cao}, \citenamefont {Hallam}, \citenamefont {Lockhart}, \citenamefont {Stojevic}, \citenamefont {Green},\ and\ \citenamefont {Severini}}]{Grant2018}%
  \BibitemOpen
  \bibfield  {author} {\bibinfo {author} {\bibfnamefont {E.}~\bibnamefont {Grant}}, \bibinfo {author} {\bibfnamefont {M.}~\bibnamefont {Benedetti}}, \bibinfo {author} {\bibfnamefont {S.}~\bibnamefont {Cao}}, \bibinfo {author} {\bibfnamefont {A.}~\bibnamefont {Hallam}}, \bibinfo {author} {\bibfnamefont {J.}~\bibnamefont {Lockhart}}, \bibinfo {author} {\bibfnamefont {V.}~\bibnamefont {Stojevic}}, \bibinfo {author} {\bibfnamefont {A.~G.}\ \bibnamefont {Green}},\ and\ \bibinfo {author} {\bibfnamefont {S.}~\bibnamefont {Severini}},\ }\bibfield  {title} {\bibinfo {title} {Hierarchical quantum classifiers},\ }\href {https://doi.org/10.1038/s41534-018-0116-9} {\bibfield  {journal} {\bibinfo  {journal} {Npj Quantum Inf.}\ }\textbf {\bibinfo {volume} {4}},\ \bibinfo {pages} {65} (\bibinfo {year} {2018})}\BibitemShut {NoStop}%
\bibitem [{\citenamefont {Lubasch}\ \emph {et~al.}(2020)\citenamefont {Lubasch}, \citenamefont {Joo}, \citenamefont {Moinier}, \citenamefont {Kiffner},\ and\ \citenamefont {Jaksch}}]{PhysRevA.101.010301}%
  \BibitemOpen
  \bibfield  {author} {\bibinfo {author} {\bibfnamefont {M.}~\bibnamefont {Lubasch}}, \bibinfo {author} {\bibfnamefont {J.}~\bibnamefont {Joo}}, \bibinfo {author} {\bibfnamefont {P.}~\bibnamefont {Moinier}}, \bibinfo {author} {\bibfnamefont {M.}~\bibnamefont {Kiffner}},\ and\ \bibinfo {author} {\bibfnamefont {D.}~\bibnamefont {Jaksch}},\ }\bibfield  {title} {\bibinfo {title} {Variational quantum algorithms for nonlinear problems},\ }\href {https://doi.org/10.1103/PhysRevA.101.010301} {\bibfield  {journal} {\bibinfo  {journal} {Phys. Rev. A}\ }\textbf {\bibinfo {volume} {101}},\ \bibinfo {pages} {010301} (\bibinfo {year} {2020})}\BibitemShut {NoStop}%
\bibitem [{\citenamefont {Vidal}(2008)}]{Vidal_2008}%
  \BibitemOpen
  \bibfield  {author} {\bibinfo {author} {\bibfnamefont {G.}~\bibnamefont {Vidal}},\ }\bibfield  {title} {\bibinfo {title} {Class of quantum many-body states that can be efficiently simulated},\ }\href {https://doi.org/10.1103/PhysRevLett.101.110501} {\bibfield  {journal} {\bibinfo  {journal} {Phys. Rev. Lett.}\ }\textbf {\bibinfo {volume} {101}},\ \bibinfo {pages} {110501} (\bibinfo {year} {2008})}\BibitemShut {NoStop}%
\bibitem [{\citenamefont {Evenbly}\ and\ \citenamefont {Vidal}(2009)}]{Evenbly_Vidal_PhysRevB.79.144108}%
  \BibitemOpen
  \bibfield  {author} {\bibinfo {author} {\bibfnamefont {G.}~\bibnamefont {Evenbly}}\ and\ \bibinfo {author} {\bibfnamefont {G.}~\bibnamefont {Vidal}},\ }\bibfield  {title} {\bibinfo {title} {Algorithms for entanglement renormalization},\ }\href {https://doi.org/10.1103/PhysRevB.79.144108} {\bibfield  {journal} {\bibinfo  {journal} {Phys. Rev. B}\ }\textbf {\bibinfo {volume} {79}},\ \bibinfo {pages} {144108} (\bibinfo {year} {2009})}\BibitemShut {NoStop}%
\bibitem [{\citenamefont {Evenbly}\ and\ \citenamefont {Vidal}(2014{\natexlab{a}})}]{Evenbly_2014}%
  \BibitemOpen
  \bibfield  {author} {\bibinfo {author} {\bibfnamefont {G.}~\bibnamefont {Evenbly}}\ and\ \bibinfo {author} {\bibfnamefont {G.}~\bibnamefont {Vidal}},\ }\bibfield  {title} {\bibinfo {title} {Class of highly entangled many-body states that can be efficiently simulated},\ }\href {https://doi.org/10.1103/PhysRevLett.112.240502} {\bibfield  {journal} {\bibinfo  {journal} {Phys. Rev. Lett.}\ }\textbf {\bibinfo {volume} {112}},\ \bibinfo {pages} {240502} (\bibinfo {year} {2014}{\natexlab{a}})}\BibitemShut {NoStop}%
\bibitem [{\citenamefont {Evenbly}\ and\ \citenamefont {Vidal}(2014{\natexlab{b}})}]{PhysRevB.89.235113}%
  \BibitemOpen
  \bibfield  {author} {\bibinfo {author} {\bibfnamefont {G.}~\bibnamefont {Evenbly}}\ and\ \bibinfo {author} {\bibfnamefont {G.}~\bibnamefont {Vidal}},\ }\bibfield  {title} {\bibinfo {title} {Scaling of entanglement entropy in the (branching) multiscale entanglement renormalization ansatz},\ }\href {https://doi.org/10.1103/PhysRevB.89.235113} {\bibfield  {journal} {\bibinfo  {journal} {Phys. Rev. B}\ }\textbf {\bibinfo {volume} {89}},\ \bibinfo {pages} {235113} (\bibinfo {year} {2014}{\natexlab{b}})}\BibitemShut {NoStop}%
\bibitem [{\citenamefont {Hauru}\ \emph {et~al.}(2021)\citenamefont {Hauru}, \citenamefont {Damme},\ and\ \citenamefont {Haegeman}}]{10.21468/SciPostPhys.10.2.040}%
  \BibitemOpen
  \bibfield  {author} {\bibinfo {author} {\bibfnamefont {M.}~\bibnamefont {Hauru}}, \bibinfo {author} {\bibfnamefont {M.~V.}\ \bibnamefont {Damme}},\ and\ \bibinfo {author} {\bibfnamefont {J.}~\bibnamefont {Haegeman}},\ }\bibfield  {title} {\bibinfo {title} {{Riemannian optimization of isometric tensor networks}},\ }\href {https://doi.org/10.21468/SciPostPhys.10.2.040} {\bibfield  {journal} {\bibinfo  {journal} {SciPost Phys.}\ }\textbf {\bibinfo {volume} {10}},\ \bibinfo {pages} {040} (\bibinfo {year} {2021})}\BibitemShut {NoStop}%
\bibitem [{\citenamefont {Mitarai}\ \emph {et~al.}(2018)\citenamefont {Mitarai}, \citenamefont {Negoro}, \citenamefont {Kitagawa},\ and\ \citenamefont {Fujii}}]{Mitarai_PRA2018}%
  \BibitemOpen
  \bibfield  {author} {\bibinfo {author} {\bibfnamefont {K.}~\bibnamefont {Mitarai}}, \bibinfo {author} {\bibfnamefont {M.}~\bibnamefont {Negoro}}, \bibinfo {author} {\bibfnamefont {M.}~\bibnamefont {Kitagawa}},\ and\ \bibinfo {author} {\bibfnamefont {K.}~\bibnamefont {Fujii}},\ }\bibfield  {title} {\bibinfo {title} {Quantum circuit learning},\ }\href {https://doi.org/10.1103/PhysRevA.98.032309} {\bibfield  {journal} {\bibinfo  {journal} {Phys. Rev. A}\ }\textbf {\bibinfo {volume} {98}},\ \bibinfo {pages} {032309} (\bibinfo {year} {2018})}\BibitemShut {NoStop}%
\bibitem [{\citenamefont {Tilly}\ \emph {et~al.}(2022)\citenamefont {Tilly}, \citenamefont {Chen}, \citenamefont {Cao}, \citenamefont {Picozzi}, \citenamefont {Setia}, \citenamefont {Li}, \citenamefont {Grant}, \citenamefont {Wossnig}, \citenamefont {Rungger}, \citenamefont {Booth},\ and\ \citenamefont {Tennyson}}]{TILLY20221}%
  \BibitemOpen
  \bibfield  {author} {\bibinfo {author} {\bibfnamefont {J.}~\bibnamefont {Tilly}}, \bibinfo {author} {\bibfnamefont {H.}~\bibnamefont {Chen}}, \bibinfo {author} {\bibfnamefont {S.}~\bibnamefont {Cao}}, \bibinfo {author} {\bibfnamefont {D.}~\bibnamefont {Picozzi}}, \bibinfo {author} {\bibfnamefont {K.}~\bibnamefont {Setia}}, \bibinfo {author} {\bibfnamefont {Y.}~\bibnamefont {Li}}, \bibinfo {author} {\bibfnamefont {E.}~\bibnamefont {Grant}}, \bibinfo {author} {\bibfnamefont {L.}~\bibnamefont {Wossnig}}, \bibinfo {author} {\bibfnamefont {I.}~\bibnamefont {Rungger}}, \bibinfo {author} {\bibfnamefont {G.~H.}\ \bibnamefont {Booth}},\ and\ \bibinfo {author} {\bibfnamefont {J.}~\bibnamefont {Tennyson}},\ }\bibfield  {title} {\bibinfo {title} {The variational quantum eigensolver: A review of methods and best practices},\ }\href {https://doi.org/https://doi.org/10.1016/j.physrep.2022.08.003} {\bibfield  {journal} {\bibinfo  {journal} {Phys. Rep.}\ }\textbf {\bibinfo {volume} {986}},\ \bibinfo {pages} {1} (\bibinfo
  {year} {2022})}\BibitemShut {NoStop}%
\bibitem [{\citenamefont {Rudolph}\ \emph {et~al.}(2022{\natexlab{b}})\citenamefont {Rudolph}, \citenamefont {Chen}, \citenamefont {Miller}, \citenamefont {Acharya},\ and\ \citenamefont {Perdomo-Ortiz}}]{rudolph2022decomposition}%
  \BibitemOpen
  \bibfield  {author} {\bibinfo {author} {\bibfnamefont {M.~S.}\ \bibnamefont {Rudolph}}, \bibinfo {author} {\bibfnamefont {J.}~\bibnamefont {Chen}}, \bibinfo {author} {\bibfnamefont {J.}~\bibnamefont {Miller}}, \bibinfo {author} {\bibfnamefont {A.}~\bibnamefont {Acharya}},\ and\ \bibinfo {author} {\bibfnamefont {A.}~\bibnamefont {Perdomo-Ortiz}},\ }\href@noop {} {\bibinfo {title} {Decomposition of matrix product states into shallow quantum circuits}} (\bibinfo {year} {2022}{\natexlab{b}}),\ \Eprint {https://arxiv.org/abs/2209.00595} {arXiv:2209.00595 [quant-ph]} \BibitemShut {NoStop}%
\bibitem [{\citenamefont {Ran}(2020)}]{Ran_2020}%
  \BibitemOpen
  \bibfield  {author} {\bibinfo {author} {\bibfnamefont {S.-J.}\ \bibnamefont {Ran}},\ }\bibfield  {title} {\bibinfo {title} {Encoding of matrix product states into quantum circuits of one- and two-qubit gates},\ }\href {https://doi.org/10.1103/PhysRevA.101.032310} {\bibfield  {journal} {\bibinfo  {journal} {Phys. Rev. A}\ }\textbf {\bibinfo {volume} {101}},\ \bibinfo {pages} {032310} (\bibinfo {year} {2020})}\BibitemShut {NoStop}%
\bibitem [{\citenamefont {Tucci}(2005)}]{tucci2005introduction}%
  \BibitemOpen
  \bibfield  {author} {\bibinfo {author} {\bibfnamefont {R.~R.}\ \bibnamefont {Tucci}},\ }\href@noop {} {\bibinfo {title} {An introduction to cartan's kak decomposition for qc programmers}} (\bibinfo {year} {2005}),\ \Eprint {https://arxiv.org/abs/quant-ph/0507171} {arXiv:quant-ph/0507171 [quant-ph]} \BibitemShut {NoStop}%
\bibitem [{\citenamefont {Kraus}\ and\ \citenamefont {Cirac}(2001)}]{PhysRevA.63.062309}%
  \BibitemOpen
  \bibfield  {author} {\bibinfo {author} {\bibfnamefont {B.}~\bibnamefont {Kraus}}\ and\ \bibinfo {author} {\bibfnamefont {J.~I.}\ \bibnamefont {Cirac}},\ }\bibfield  {title} {\bibinfo {title} {Optimal creation of entanglement using a two-qubit gate},\ }\href {https://doi.org/10.1103/PhysRevA.63.062309} {\bibfield  {journal} {\bibinfo  {journal} {Phys. Rev. A}\ }\textbf {\bibinfo {volume} {63}},\ \bibinfo {pages} {062309} (\bibinfo {year} {2001})}\BibitemShut {NoStop}%
\bibitem [{\citenamefont {DiVincenzo}\ and\ \citenamefont {Smolin}(1994)}]{DiVincenzo_PHYCMP1994}%
  \BibitemOpen
  \bibfield  {author} {\bibinfo {author} {\bibfnamefont {D.}~\bibnamefont {DiVincenzo}}\ and\ \bibinfo {author} {\bibfnamefont {J.}~\bibnamefont {Smolin}},\ }\bibfield  {title} {\bibinfo {title} {Results on two-bit gate design for quantum computers},\ }in\ \href {https://doi.org/10.1109/PHYCMP.1994.363704} {\emph {\bibinfo {booktitle} {Proceedings Workshop on Physics and Computation. PhysComp '94}}}\ (\bibinfo {year} {1994})\ pp.\ \bibinfo {pages} {14--23}\BibitemShut {NoStop}%
\bibitem [{\citenamefont {Amy}\ \emph {et~al.}(2013)\citenamefont {Amy}, \citenamefont {Maslov}, \citenamefont {Mosca},\ and\ \citenamefont {Roetteler}}]{Amy_TCAD2013}%
  \BibitemOpen
  \bibfield  {author} {\bibinfo {author} {\bibfnamefont {M.}~\bibnamefont {Amy}}, \bibinfo {author} {\bibfnamefont {D.}~\bibnamefont {Maslov}}, \bibinfo {author} {\bibfnamefont {M.}~\bibnamefont {Mosca}},\ and\ \bibinfo {author} {\bibfnamefont {M.}~\bibnamefont {Roetteler}},\ }\bibfield  {title} {\bibinfo {title} {A meet-in-the-middle algorithm for fast synthesis of depth-optimal quantum circuits},\ }\href {https://doi.org/10.1109/TCAD.2013.2244643} {\bibfield  {journal} {\bibinfo  {journal} {IEEE T. Comput. Aid. D.}\ }\textbf {\bibinfo {volume} {32}},\ \bibinfo {pages} {818} (\bibinfo {year} {2013})}\BibitemShut {NoStop}%
\bibitem [{\citenamefont {Nam}\ \emph {et~al.}(2018)\citenamefont {Nam}, \citenamefont {Ross}, \citenamefont {Su}, \citenamefont {Childs},\ and\ \citenamefont {Maslov}}]{Nam_npjqi2018}%
  \BibitemOpen
  \bibfield  {author} {\bibinfo {author} {\bibfnamefont {Y.}~\bibnamefont {Nam}}, \bibinfo {author} {\bibfnamefont {N.~J.}\ \bibnamefont {Ross}}, \bibinfo {author} {\bibfnamefont {Y.}~\bibnamefont {Su}}, \bibinfo {author} {\bibfnamefont {A.~M.}\ \bibnamefont {Childs}},\ and\ \bibinfo {author} {\bibfnamefont {D.}~\bibnamefont {Maslov}},\ }\bibfield  {title} {\bibinfo {title} {Automated optimization of large quantum circuits with continuous parameters},\ }\href {https://doi.org/10.1038/s41534-018-0072-4} {\bibfield  {journal} {\bibinfo  {journal} {Npj Quantum Inf.}\ }\textbf {\bibinfo {volume} {4}},\ \bibinfo {pages} {23} (\bibinfo {year} {2018})}\BibitemShut {NoStop}%
\bibitem [{\citenamefont {Khatri}\ \emph {et~al.}(2019)\citenamefont {Khatri}, \citenamefont {LaRose}, \citenamefont {Poremba}, \citenamefont {Cincio}, \citenamefont {Sornborger},\ and\ \citenamefont {Coles}}]{Khatri_Q2019}%
  \BibitemOpen
  \bibfield  {author} {\bibinfo {author} {\bibfnamefont {S.}~\bibnamefont {Khatri}}, \bibinfo {author} {\bibfnamefont {R.}~\bibnamefont {LaRose}}, \bibinfo {author} {\bibfnamefont {A.}~\bibnamefont {Poremba}}, \bibinfo {author} {\bibfnamefont {L.}~\bibnamefont {Cincio}}, \bibinfo {author} {\bibfnamefont {A.~T.}\ \bibnamefont {Sornborger}},\ and\ \bibinfo {author} {\bibfnamefont {P.~J.}\ \bibnamefont {Coles}},\ }\bibfield  {title} {\bibinfo {title} {Quantum-assisted quantum compiling},\ }\href {https://doi.org/10.22331/q-2019-05-13-140} {\bibfield  {journal} {\bibinfo  {journal} {{Quantum}}\ }\textbf {\bibinfo {volume} {3}},\ \bibinfo {pages} {140} (\bibinfo {year} {2019})}\BibitemShut {NoStop}%
\bibitem [{\citenamefont {Nagarajan}\ \emph {et~al.}(2021)\citenamefont {Nagarajan}, \citenamefont {Lockwood},\ and\ \citenamefont {Coffrin}}]{Nagarajan_QCS2021}%
  \BibitemOpen
  \bibfield  {author} {\bibinfo {author} {\bibfnamefont {H.}~\bibnamefont {Nagarajan}}, \bibinfo {author} {\bibfnamefont {O.}~\bibnamefont {Lockwood}},\ and\ \bibinfo {author} {\bibfnamefont {C.}~\bibnamefont {Coffrin}},\ }\bibfield  {title} {\bibinfo {title} {Quantumcircuitopt: An open-source framework for provably optimal quantum circuit design},\ }in\ \href {https://doi.org/10.1109/QCS54837.2021.00010} {\emph {\bibinfo {booktitle} {2021 IEEE/ACM Second International Workshop on Quantum Computing Software (QCS)}}}\ (\bibinfo {year} {2021})\ pp.\ \bibinfo {pages} {55--63}\BibitemShut {NoStop}%
\bibitem [{\citenamefont {Younis}\ \emph {et~al.}(2021)\citenamefont {Younis}, \citenamefont {Sen}, \citenamefont {Yelick},\ and\ \citenamefont {Iancu}}]{younis2021qfast}%
  \BibitemOpen
  \bibfield  {author} {\bibinfo {author} {\bibfnamefont {E.}~\bibnamefont {Younis}}, \bibinfo {author} {\bibfnamefont {K.}~\bibnamefont {Sen}}, \bibinfo {author} {\bibfnamefont {K.}~\bibnamefont {Yelick}},\ and\ \bibinfo {author} {\bibfnamefont {C.}~\bibnamefont {Iancu}},\ }\href@noop {} {\bibinfo {title} {Qfast: Conflating search and numerical optimization for scalable quantum circuit synthesis}} (\bibinfo {year} {2021}),\ \Eprint {https://arxiv.org/abs/2103.07093} {arXiv:2103.07093 [quant-ph]} \BibitemShut {NoStop}%
\bibitem [{\citenamefont {Fösel}\ \emph {et~al.}(2021)\citenamefont {Fösel}, \citenamefont {Niu}, \citenamefont {Marquardt},\ and\ \citenamefont {Li}}]{fosel2021quantum}%
  \BibitemOpen
  \bibfield  {author} {\bibinfo {author} {\bibfnamefont {T.}~\bibnamefont {Fösel}}, \bibinfo {author} {\bibfnamefont {M.~Y.}\ \bibnamefont {Niu}}, \bibinfo {author} {\bibfnamefont {F.}~\bibnamefont {Marquardt}},\ and\ \bibinfo {author} {\bibfnamefont {L.}~\bibnamefont {Li}},\ }\href@noop {} {\bibinfo {title} {Quantum circuit optimization with deep reinforcement learning}} (\bibinfo {year} {2021}),\ \Eprint {https://arxiv.org/abs/2103.07585} {arXiv:2103.07585 [quant-ph]} \BibitemShut {NoStop}%
\bibitem [{\citenamefont {Shirakawa}\ \emph {et~al.}(2021)\citenamefont {Shirakawa}, \citenamefont {Ueda},\ and\ \citenamefont {Yunoki}}]{AQCE}%
  \BibitemOpen
  \bibfield  {author} {\bibinfo {author} {\bibfnamefont {T.}~\bibnamefont {Shirakawa}}, \bibinfo {author} {\bibfnamefont {H.}~\bibnamefont {Ueda}},\ and\ \bibinfo {author} {\bibfnamefont {S.}~\bibnamefont {Yunoki}},\ }\href@noop {} {\bibinfo {title} {Automatic quantum circuit encoding of a given arbitrary quantum state}} (\bibinfo {year} {2021}),\ \Eprint {https://arxiv.org/abs/2112.14524} {arXiv:2112.14524 [quant-ph]} \BibitemShut {NoStop}%
\bibitem [{\citenamefont {Meister}\ \emph {et~al.}(2022)\citenamefont {Meister}, \citenamefont {Gustiani},\ and\ \citenamefont {Benjamin}}]{meister2022exploring}%
  \BibitemOpen
  \bibfield  {author} {\bibinfo {author} {\bibfnamefont {R.}~\bibnamefont {Meister}}, \bibinfo {author} {\bibfnamefont {C.}~\bibnamefont {Gustiani}},\ and\ \bibinfo {author} {\bibfnamefont {S.~C.}\ \bibnamefont {Benjamin}},\ }\href@noop {} {\bibinfo {title} {Exploring ab initio machine synthesis of quantum circuits}} (\bibinfo {year} {2022}),\ \Eprint {https://arxiv.org/abs/2206.11245} {arXiv:2206.11245 [quant-ph]} \BibitemShut {NoStop}%
\bibitem [{\citenamefont {Rakyta}\ and\ \citenamefont {Zimborás}(2022)}]{rakyta2022efficient}%
  \BibitemOpen
  \bibfield  {author} {\bibinfo {author} {\bibfnamefont {P.}~\bibnamefont {Rakyta}}\ and\ \bibinfo {author} {\bibfnamefont {Z.}~\bibnamefont {Zimborás}},\ }\href@noop {} {\bibinfo {title} {Efficient quantum gate decomposition via adaptive circuit compression}} (\bibinfo {year} {2022}),\ \Eprint {https://arxiv.org/abs/2203.04426} {arXiv:2203.04426 [quant-ph]} \BibitemShut {NoStop}%
\bibitem [{\citenamefont {Nemkov}\ \emph {et~al.}(2023)\citenamefont {Nemkov}, \citenamefont {Kiktenko}, \citenamefont {Luchnikov},\ and\ \citenamefont {Fedorov}}]{nemkov2023efficient}%
  \BibitemOpen
  \bibfield  {author} {\bibinfo {author} {\bibfnamefont {N.~A.}\ \bibnamefont {Nemkov}}, \bibinfo {author} {\bibfnamefont {E.~O.}\ \bibnamefont {Kiktenko}}, \bibinfo {author} {\bibfnamefont {I.~A.}\ \bibnamefont {Luchnikov}},\ and\ \bibinfo {author} {\bibfnamefont {A.~K.}\ \bibnamefont {Fedorov}},\ }\href@noop {} {\bibinfo {title} {Efficient variational synthesis of quantum circuits with coherent multi-start optimization}} (\bibinfo {year} {2023}),\ \Eprint {https://arxiv.org/abs/2205.01121} {arXiv:2205.01121 [quant-ph]} \BibitemShut {NoStop}%
\bibitem [{\citenamefont {Smith}\ \emph {et~al.}(2023)\citenamefont {Smith}, \citenamefont {Davis}, \citenamefont {Larson}, \citenamefont {Younis}, \citenamefont {Oftelie}, \citenamefont {Lavrijsen},\ and\ \citenamefont {Iancu}}]{Smith_2023}%
  \BibitemOpen
  \bibfield  {author} {\bibinfo {author} {\bibfnamefont {E.}~\bibnamefont {Smith}}, \bibinfo {author} {\bibfnamefont {M.~G.}\ \bibnamefont {Davis}}, \bibinfo {author} {\bibfnamefont {J.}~\bibnamefont {Larson}}, \bibinfo {author} {\bibfnamefont {E.}~\bibnamefont {Younis}}, \bibinfo {author} {\bibfnamefont {L.~B.}\ \bibnamefont {Oftelie}}, \bibinfo {author} {\bibfnamefont {W.}~\bibnamefont {Lavrijsen}},\ and\ \bibinfo {author} {\bibfnamefont {C.}~\bibnamefont {Iancu}},\ }\bibfield  {title} {\bibinfo {title} {Leap: Scaling numerical optimization based synthesis using an incremental approach},\ }\href {https://doi.org/10.1145/3548693} {\bibfield  {journal} {\bibinfo  {journal} {ACM Trans. Quantum Comput.}\ }\textbf {\bibinfo {volume} {4}},\ \bibinfo {pages} {5} (\bibinfo {year} {2023})}\BibitemShut {NoStop}%
\bibitem [{\citenamefont {Shi}\ \emph {et~al.}(2006)\citenamefont {Shi}, \citenamefont {Duan},\ and\ \citenamefont {Vidal}}]{PhysRevA.74.022320}%
  \BibitemOpen
  \bibfield  {author} {\bibinfo {author} {\bibfnamefont {Y.-Y.}\ \bibnamefont {Shi}}, \bibinfo {author} {\bibfnamefont {L.-M.}\ \bibnamefont {Duan}},\ and\ \bibinfo {author} {\bibfnamefont {G.}~\bibnamefont {Vidal}},\ }\bibfield  {title} {\bibinfo {title} {Classical simulation of quantum many-body systems with a tree tensor network},\ }\href {https://doi.org/10.1103/PhysRevA.74.022320} {\bibfield  {journal} {\bibinfo  {journal} {Phys. Rev. A}\ }\textbf {\bibinfo {volume} {74}},\ \bibinfo {pages} {022320} (\bibinfo {year} {2006})}\BibitemShut {NoStop}%
\bibitem [{\citenamefont {Tagliacozzo}\ \emph {et~al.}(2009)\citenamefont {Tagliacozzo}, \citenamefont {Evenbly},\ and\ \citenamefont {Vidal}}]{PhysRevB.80.235127}%
  \BibitemOpen
  \bibfield  {author} {\bibinfo {author} {\bibfnamefont {L.}~\bibnamefont {Tagliacozzo}}, \bibinfo {author} {\bibfnamefont {G.}~\bibnamefont {Evenbly}},\ and\ \bibinfo {author} {\bibfnamefont {G.}~\bibnamefont {Vidal}},\ }\bibfield  {title} {\bibinfo {title} {Simulation of two-dimensional quantum systems using a tree tensor network that exploits the entropic area law},\ }\href {https://doi.org/10.1103/PhysRevB.80.235127} {\bibfield  {journal} {\bibinfo  {journal} {Phys. Rev. B}\ }\textbf {\bibinfo {volume} {80}},\ \bibinfo {pages} {235127} (\bibinfo {year} {2009})}\BibitemShut {NoStop}%
\bibitem [{\citenamefont {Murg}\ \emph {et~al.}(2010)\citenamefont {Murg}, \citenamefont {Verstraete}, \citenamefont {Legeza},\ and\ \citenamefont {Noack}}]{PhysRevB.82.205105}%
  \BibitemOpen
  \bibfield  {author} {\bibinfo {author} {\bibfnamefont {V.}~\bibnamefont {Murg}}, \bibinfo {author} {\bibfnamefont {F.}~\bibnamefont {Verstraete}}, \bibinfo {author} {\bibfnamefont {O.}~\bibnamefont {Legeza}},\ and\ \bibinfo {author} {\bibfnamefont {R.~M.}\ \bibnamefont {Noack}},\ }\bibfield  {title} {\bibinfo {title} {Simulating strongly correlated quantum systems with tree tensor networks},\ }\href {https://doi.org/10.1103/PhysRevB.82.205105} {\bibfield  {journal} {\bibinfo  {journal} {Phys. Rev. B}\ }\textbf {\bibinfo {volume} {82}},\ \bibinfo {pages} {205105} (\bibinfo {year} {2010})}\BibitemShut {NoStop}%
\bibitem [{\citenamefont {Fishman}\ \emph {et~al.}(2022)\citenamefont {Fishman}, \citenamefont {White},\ and\ \citenamefont {Stoudenmire}}]{itensor}%
  \BibitemOpen
  \bibfield  {author} {\bibinfo {author} {\bibfnamefont {M.}~\bibnamefont {Fishman}}, \bibinfo {author} {\bibfnamefont {S.~R.}\ \bibnamefont {White}},\ and\ \bibinfo {author} {\bibfnamefont {E.~M.}\ \bibnamefont {Stoudenmire}},\ }\bibfield  {title} {\bibinfo {title} {{The ITensor Software Library for Tensor Network Calculations}},\ }\href {https://doi.org/10.21468/SciPostPhysCodeb.4} {\bibfield  {journal} {\bibinfo  {journal} {SciPost Phys. Codebases}\ ,\ \bibinfo {pages} {4}} (\bibinfo {year} {2022})}\BibitemShut {NoStop}%
\bibitem [{\citenamefont {Developers}(2023)}]{cirq}%
  \BibitemOpen
  \bibfield  {author} {\bibinfo {author} {\bibfnamefont {C.}~\bibnamefont {Developers}},\ }\href {https://doi.org/10.5281/zenodo.10247207} {\bibinfo {title} {Cirq}} (\bibinfo {year} {2023})\BibitemShut {NoStop}%
\bibitem [{\citenamefont {Suzuki}\ \emph {et~al.}(2021)\citenamefont {Suzuki}, \citenamefont {Kawase}, \citenamefont {Masumura}, \citenamefont {Hiraga}, \citenamefont {Nakadai}, \citenamefont {Chen}, \citenamefont {Nakanishi}, \citenamefont {Mitarai}, \citenamefont {Imai}, \citenamefont {Tamiya}, \citenamefont {Yamamoto}, \citenamefont {Yan}, \citenamefont {Kawakubo}, \citenamefont {Nakagawa}, \citenamefont {Ibe}, \citenamefont {Zhang}, \citenamefont {Yamashita}, \citenamefont {Yoshimura}, \citenamefont {Hayashi},\ and\ \citenamefont {Fujii}}]{Suzuki2021qulacsfast}%
  \BibitemOpen
  \bibfield  {author} {\bibinfo {author} {\bibfnamefont {Y.}~\bibnamefont {Suzuki}}, \bibinfo {author} {\bibfnamefont {Y.}~\bibnamefont {Kawase}}, \bibinfo {author} {\bibfnamefont {Y.}~\bibnamefont {Masumura}}, \bibinfo {author} {\bibfnamefont {Y.}~\bibnamefont {Hiraga}}, \bibinfo {author} {\bibfnamefont {M.}~\bibnamefont {Nakadai}}, \bibinfo {author} {\bibfnamefont {J.}~\bibnamefont {Chen}}, \bibinfo {author} {\bibfnamefont {K.~M.}\ \bibnamefont {Nakanishi}}, \bibinfo {author} {\bibfnamefont {K.}~\bibnamefont {Mitarai}}, \bibinfo {author} {\bibfnamefont {R.}~\bibnamefont {Imai}}, \bibinfo {author} {\bibfnamefont {S.}~\bibnamefont {Tamiya}}, \bibinfo {author} {\bibfnamefont {T.}~\bibnamefont {Yamamoto}}, \bibinfo {author} {\bibfnamefont {T.}~\bibnamefont {Yan}}, \bibinfo {author} {\bibfnamefont {T.}~\bibnamefont {Kawakubo}}, \bibinfo {author} {\bibfnamefont {Y.~O.}\ \bibnamefont {Nakagawa}}, \bibinfo {author} {\bibfnamefont {Y.}~\bibnamefont {Ibe}}, \bibinfo {author} {\bibfnamefont {Y.}~\bibnamefont {Zhang}},
  \bibinfo {author} {\bibfnamefont {H.}~\bibnamefont {Yamashita}}, \bibinfo {author} {\bibfnamefont {H.}~\bibnamefont {Yoshimura}}, \bibinfo {author} {\bibfnamefont {A.}~\bibnamefont {Hayashi}},\ and\ \bibinfo {author} {\bibfnamefont {K.}~\bibnamefont {Fujii}},\ }\bibfield  {title} {\bibinfo {title} {Qulacs: a fast and versatile quantum circuit simulator for research purpose},\ }\href {https://doi.org/10.22331/q-2021-10-06-559} {\bibfield  {journal} {\bibinfo  {journal} {{Quantum}}\ }\textbf {\bibinfo {volume} {5}},\ \bibinfo {pages} {559} (\bibinfo {year} {2021})}\BibitemShut {NoStop}%
\bibitem [{\citenamefont {Virtanen}\ \emph {et~al.}(2020)\citenamefont {Virtanen}, \citenamefont {Gommers}, \citenamefont {Oliphant}, \citenamefont {Haberland}, \citenamefont {Reddy}, \citenamefont {Cournapeau}, \citenamefont {Burovski}, \citenamefont {Peterson}, \citenamefont {Weckesser}, \citenamefont {Bright}, \citenamefont {{van der Walt}}, \citenamefont {Brett}, \citenamefont {Wilson}, \citenamefont {Millman}, \citenamefont {Mayorov}, \citenamefont {Nelson}, \citenamefont {Jones}, \citenamefont {Kern}, \citenamefont {Larson}, \citenamefont {Carey}, \citenamefont {Polat}, \citenamefont {Feng}, \citenamefont {Moore}, \citenamefont {{VanderPlas}}, \citenamefont {Laxalde}, \citenamefont {Perktold}, \citenamefont {Cimrman}, \citenamefont {Henriksen}, \citenamefont {Quintero}, \citenamefont {Harris}, \citenamefont {Archibald}, \citenamefont {Ribeiro}, \citenamefont {Pedregosa}, \citenamefont {{van Mulbregt}},\ and\ \citenamefont {{SciPy 1.0 Contributors}}}]{2020SciPy-NMeth}%
  \BibitemOpen
  \bibfield  {author} {\bibinfo {author} {\bibfnamefont {P.}~\bibnamefont {Virtanen}}, \bibinfo {author} {\bibfnamefont {R.}~\bibnamefont {Gommers}}, \bibinfo {author} {\bibfnamefont {T.~E.}\ \bibnamefont {Oliphant}}, \bibinfo {author} {\bibfnamefont {M.}~\bibnamefont {Haberland}}, \bibinfo {author} {\bibfnamefont {T.}~\bibnamefont {Reddy}}, \bibinfo {author} {\bibfnamefont {D.}~\bibnamefont {Cournapeau}}, \bibinfo {author} {\bibfnamefont {E.}~\bibnamefont {Burovski}}, \bibinfo {author} {\bibfnamefont {P.}~\bibnamefont {Peterson}}, \bibinfo {author} {\bibfnamefont {W.}~\bibnamefont {Weckesser}}, \bibinfo {author} {\bibfnamefont {J.}~\bibnamefont {Bright}}, \bibinfo {author} {\bibfnamefont {S.~J.}\ \bibnamefont {{van der Walt}}}, \bibinfo {author} {\bibfnamefont {M.}~\bibnamefont {Brett}}, \bibinfo {author} {\bibfnamefont {J.}~\bibnamefont {Wilson}}, \bibinfo {author} {\bibfnamefont {K.~J.}\ \bibnamefont {Millman}}, \bibinfo {author} {\bibfnamefont {N.}~\bibnamefont {Mayorov}}, \bibinfo {author} {\bibfnamefont
  {A.~R.~J.}\ \bibnamefont {Nelson}}, \bibinfo {author} {\bibfnamefont {E.}~\bibnamefont {Jones}}, \bibinfo {author} {\bibfnamefont {R.}~\bibnamefont {Kern}}, \bibinfo {author} {\bibfnamefont {E.}~\bibnamefont {Larson}}, \bibinfo {author} {\bibfnamefont {C.~J.}\ \bibnamefont {Carey}}, \bibinfo {author} {\bibfnamefont {{\.I}.}~\bibnamefont {Polat}}, \bibinfo {author} {\bibfnamefont {Y.}~\bibnamefont {Feng}}, \bibinfo {author} {\bibfnamefont {E.~W.}\ \bibnamefont {Moore}}, \bibinfo {author} {\bibfnamefont {J.}~\bibnamefont {{VanderPlas}}}, \bibinfo {author} {\bibfnamefont {D.}~\bibnamefont {Laxalde}}, \bibinfo {author} {\bibfnamefont {J.}~\bibnamefont {Perktold}}, \bibinfo {author} {\bibfnamefont {R.}~\bibnamefont {Cimrman}}, \bibinfo {author} {\bibfnamefont {I.}~\bibnamefont {Henriksen}}, \bibinfo {author} {\bibfnamefont {E.~A.}\ \bibnamefont {Quintero}}, \bibinfo {author} {\bibfnamefont {C.~R.}\ \bibnamefont {Harris}}, \bibinfo {author} {\bibfnamefont {A.~M.}\ \bibnamefont {Archibald}}, \bibinfo {author}
  {\bibfnamefont {A.~H.}\ \bibnamefont {Ribeiro}}, \bibinfo {author} {\bibfnamefont {F.}~\bibnamefont {Pedregosa}}, \bibinfo {author} {\bibfnamefont {P.}~\bibnamefont {{van Mulbregt}}},\ and\ \bibinfo {author} {\bibnamefont {{SciPy 1.0 Contributors}}},\ }\bibfield  {title} {\bibinfo {title} {{{SciPy} 1.0: Fundamental Algorithms for Scientific Computing in Python}},\ }\href {https://doi.org/10.1038/s41592-019-0686-2} {\bibfield  {journal} {\bibinfo  {journal} {Nat. Methods}\ }\textbf {\bibinfo {volume} {17}},\ \bibinfo {pages} {261} (\bibinfo {year} {2020})}\BibitemShut {NoStop}%
\bibitem [{\citenamefont {Li}\ \emph {et~al.}(2022)\citenamefont {Li}, \citenamefont {Ren}, \citenamefont {Yang},\ and\ \citenamefont {Shuai}}]{Li_2022}%
  \BibitemOpen
  \bibfield  {author} {\bibinfo {author} {\bibfnamefont {W.}~\bibnamefont {Li}}, \bibinfo {author} {\bibfnamefont {J.}~\bibnamefont {Ren}}, \bibinfo {author} {\bibfnamefont {H.}~\bibnamefont {Yang}},\ and\ \bibinfo {author} {\bibfnamefont {Z.}~\bibnamefont {Shuai}},\ }\bibfield  {title} {\bibinfo {title} {On the fly swapping algorithm for ordering of degrees of freedom in density matrix renormalization group},\ }\href {https://doi.org/10.1088/1361-648X/ac640e} {\bibfield  {journal} {\bibinfo  {journal} {J. Phys. Condens. Matter}\ }\textbf {\bibinfo {volume} {34}},\ \bibinfo {pages} {254003} (\bibinfo {year} {2022})}\BibitemShut {NoStop}%
\bibitem [{\citenamefont {Hikihara}\ \emph {et~al.}(2023)\citenamefont {Hikihara}, \citenamefont {Ueda}, \citenamefont {Okunishi}, \citenamefont {Harada},\ and\ \citenamefont {Nishino}}]{HikiharaPRR2023}%
  \BibitemOpen
  \bibfield  {author} {\bibinfo {author} {\bibfnamefont {T.}~\bibnamefont {Hikihara}}, \bibinfo {author} {\bibfnamefont {H.}~\bibnamefont {Ueda}}, \bibinfo {author} {\bibfnamefont {K.}~\bibnamefont {Okunishi}}, \bibinfo {author} {\bibfnamefont {K.}~\bibnamefont {Harada}},\ and\ \bibinfo {author} {\bibfnamefont {T.}~\bibnamefont {Nishino}},\ }\bibfield  {title} {\bibinfo {title} {Automatic structural optimization of tree tensor networks},\ }\href {https://doi.org/10.1103/PhysRevResearch.5.013031} {\bibfield  {journal} {\bibinfo  {journal} {Phys. Rev. Res.}\ }\textbf {\bibinfo {volume} {5}},\ \bibinfo {pages} {013031} (\bibinfo {year} {2023})}\BibitemShut {NoStop}%
\bibitem [{\citenamefont {Okunishi}\ \emph {et~al.}(2023)\citenamefont {Okunishi}, \citenamefont {Ueda},\ and\ \citenamefont {Nishino}}]{okunishi2023ptep}%
  \BibitemOpen
  \bibfield  {author} {\bibinfo {author} {\bibfnamefont {K.}~\bibnamefont {Okunishi}}, \bibinfo {author} {\bibfnamefont {H.}~\bibnamefont {Ueda}},\ and\ \bibinfo {author} {\bibfnamefont {T.}~\bibnamefont {Nishino}},\ }\bibfield  {title} {\bibinfo {title} {Entanglement bipartitioning and tree tensor networks},\ }\href {https://doi.org/10.1093/ptep/ptad018} {\bibfield  {journal} {\bibinfo  {journal} {Prog. Theor. Exp. Phys.}\ }\textbf {\bibinfo {volume} {2023}},\ \bibinfo {pages} {023A02} (\bibinfo {year} {2023})}\BibitemShut {NoStop}%
\bibitem [{\citenamefont {Kim}\ and\ \citenamefont {Swingle}(2017)}]{DMERA}%
  \BibitemOpen
  \bibfield  {author} {\bibinfo {author} {\bibfnamefont {I.~H.}\ \bibnamefont {Kim}}\ and\ \bibinfo {author} {\bibfnamefont {B.}~\bibnamefont {Swingle}},\ }\href@noop {} {\bibinfo {title} {Robust entanglement renormalization on a noisy quantum computer}} (\bibinfo {year} {2017}),\ \Eprint {https://arxiv.org/abs/1711.07500} {arXiv:1711.07500 [quant-ph]} \BibitemShut {NoStop}%
\bibitem [{\citenamefont {Geng}\ \emph {et~al.}(2022)\citenamefont {Geng}, \citenamefont {Hu},\ and\ \citenamefont {Zou}}]{Geng_2022}%
  \BibitemOpen
  \bibfield  {author} {\bibinfo {author} {\bibfnamefont {C.}~\bibnamefont {Geng}}, \bibinfo {author} {\bibfnamefont {H.-Y.}\ \bibnamefont {Hu}},\ and\ \bibinfo {author} {\bibfnamefont {Y.}~\bibnamefont {Zou}},\ }\bibfield  {title} {\bibinfo {title} {Differentiable programming of isometric tensor networks},\ }\href {https://doi.org/10.1088/2632-2153/ac48a2} {\bibfield  {journal} {\bibinfo  {journal} {Mach. learn.: sci. technol.}\ }\textbf {\bibinfo {volume} {3}},\ \bibinfo {pages} {015020} (\bibinfo {year} {2022})}\BibitemShut {NoStop}%
\bibitem [{\citenamefont {Cong}\ \emph {et~al.}(2019)\citenamefont {Cong}, \citenamefont {Choi},\ and\ \citenamefont {Lukin}}]{Cong2019}%
  \BibitemOpen
  \bibfield  {author} {\bibinfo {author} {\bibfnamefont {I.}~\bibnamefont {Cong}}, \bibinfo {author} {\bibfnamefont {S.}~\bibnamefont {Choi}},\ and\ \bibinfo {author} {\bibfnamefont {M.~D.}\ \bibnamefont {Lukin}},\ }\bibfield  {title} {\bibinfo {title} {Quantum convolutional neural networks},\ }\href {https://doi.org/10.1038/s41567-019-0648-8} {\bibfield  {journal} {\bibinfo  {journal} {Nat Phys.}\ }\textbf {\bibinfo {volume} {15}},\ \bibinfo {pages} {1273} (\bibinfo {year} {2019})}\BibitemShut {NoStop}%
\bibitem [{\citenamefont {Lin}\ \emph {et~al.}(2021)\citenamefont {Lin}, \citenamefont {Dilip}, \citenamefont {Green}, \citenamefont {Smith},\ and\ \citenamefont {Pollmann}}]{PRXQuantum.2.010342}%
  \BibitemOpen
  \bibfield  {author} {\bibinfo {author} {\bibfnamefont {S.-H.}\ \bibnamefont {Lin}}, \bibinfo {author} {\bibfnamefont {R.}~\bibnamefont {Dilip}}, \bibinfo {author} {\bibfnamefont {A.~G.}\ \bibnamefont {Green}}, \bibinfo {author} {\bibfnamefont {A.}~\bibnamefont {Smith}},\ and\ \bibinfo {author} {\bibfnamefont {F.}~\bibnamefont {Pollmann}},\ }\bibfield  {title} {\bibinfo {title} {Real- and imaginary-time evolution with compressed quantum circuits},\ }\href {https://doi.org/10.1103/PRXQuantum.2.010342} {\bibfield  {journal} {\bibinfo  {journal} {PRX Quantum}\ }\textbf {\bibinfo {volume} {2}},\ \bibinfo {pages} {010342} (\bibinfo {year} {2021})}\BibitemShut {NoStop}%
\bibitem [{\citenamefont {Vitagliano}\ \emph {et~al.}(2010)\citenamefont {Vitagliano}, \citenamefont {Riera},\ and\ \citenamefont {Latorre}}]{Vitagliano_2010}%
  \BibitemOpen
  \bibfield  {author} {\bibinfo {author} {\bibfnamefont {G.}~\bibnamefont {Vitagliano}}, \bibinfo {author} {\bibfnamefont {A.}~\bibnamefont {Riera}},\ and\ \bibinfo {author} {\bibfnamefont {J.~I.}\ \bibnamefont {Latorre}},\ }\bibfield  {title} {\bibinfo {title} {Volume-law scaling for the entanglement entropy in spin-1/2 chains},\ }\href {https://doi.org/10.1088/1367-2630/12/11/113049} {\bibfield  {journal} {\bibinfo  {journal} {New J. Phys.}\ }\textbf {\bibinfo {volume} {12}},\ \bibinfo {pages} {113049} (\bibinfo {year} {2010})}\BibitemShut {NoStop}%
\bibitem [{\citenamefont {Ramírez}\ \emph {et~al.}(2014)\citenamefont {Ramírez}, \citenamefont {Rodríguez-Laguna},\ and\ \citenamefont {Sierra}}]{Ramirez_2014}%
  \BibitemOpen
  \bibfield  {author} {\bibinfo {author} {\bibfnamefont {G.}~\bibnamefont {Ramírez}}, \bibinfo {author} {\bibfnamefont {J.}~\bibnamefont {Rodríguez-Laguna}},\ and\ \bibinfo {author} {\bibfnamefont {G.}~\bibnamefont {Sierra}},\ }\bibfield  {title} {\bibinfo {title} {From conformal to volume law for the entanglement entropy in exponentially deformed critical spin 1/2 chains},\ }\href {https://doi.org/10.1088/1742-5468/2014/10/P10004} {\bibfield  {journal} {\bibinfo  {journal} {J. Stat. Mech.: Theory Exp.}\ }\textbf {\bibinfo {volume} {2014}}\bibinfo  {number} { (10)},\ \bibinfo {pages} {P10004}}\BibitemShut {NoStop}%
\bibitem [{\citenamefont {Ramírez}\ \emph {et~al.}(2015)\citenamefont {Ramírez}, \citenamefont {Rodríguez-Laguna},\ and\ \citenamefont {Sierra}}]{Ramirez_2015}%
  \BibitemOpen
\bibfield  {number} {  }\bibfield  {author} {\bibinfo {author} {\bibfnamefont {G.}~\bibnamefont {Ramírez}}, \bibinfo {author} {\bibfnamefont {J.}~\bibnamefont {Rodríguez-Laguna}},\ and\ \bibinfo {author} {\bibfnamefont {G.}~\bibnamefont {Sierra}},\ }\bibfield  {title} {\bibinfo {title} {Entanglement over the rainbow},\ }\href {https://doi.org/10.1088/1742-5468/2015/06/P06002} {\bibfield  {journal} {\bibinfo  {journal} {J. Stat. Mech.: Theory Exp.}\ }\textbf {\bibinfo {volume} {2015}}\bibinfo  {number} { (6)},\ \bibinfo {pages} {P06002}}\BibitemShut {NoStop}%
\end{thebibliography}%
\end{document}